\documentclass[aps,prc,twocolumn,showpacs,floatfix,nofootinbib]{revtex4}
\usepackage{graphicx}
\usepackage{dcolumn}
\usepackage{color}

\begin{document}


\title{Lithium isotopes within the {\it  ab initio} no-core full
configuration approach}


\author{Chase Cockrell}
\email[]{cockrell@iastate.edu}
\author{James P. Vary}
\email[]{jpvary@iastate.edu}
\author{Pieter Maris}
\email[]{pmaris@iastate.edu}
\affiliation{Department of Physics and Astronomy,
Iowa State University, Ames, IA 50011}

\date{\today}

\begin{abstract}
We perform no-core full configuration calculations for the Lithium
isotopes, $^6$Li, $^7$Li, and $^8$Li with the realistic
nucleon-nucleon interaction JISP16.  We obtain a set of observables,
such as spectra, radii, multipole moments, transition probabilities, etc., and
compare with experiment where available.  We also present one-body
density distributions for selected states.  Convergence properties of
these density distributions shed light on the convergence properties
of one-body observables.  We obtain underbinding by 0.5 MeV, 0.7 MeV,
and 1.0 MeV for $^6$Li, $^7$Li, $^8$Li respectively.  Magnetic moments
are well-converged and agree with experiment to within 20$\%$.

\end{abstract}



\maketitle

\section{Introduction and Motivation}

The rapid development of {\it ab initio} methods for solving finite
nuclei has opened a range of nuclear phenomena that can be evaluated
to high precision using realistic inter-nucleon interactions.  Here we
investigate three Li isotopes through direct solution of the nuclear
many-body problem with the JISP16 nucleon-nucleon (NN)
interaction~\cite{Shirokov:2005bk,Shirokov:2004ff,Shirokov:2003kk} by
diagonalization in a sufficiently large basis space that converged
binding energies are accessed by simple extrapolation.  The many-body
approach we adopt is referred to as the no-core full configuration
(NCFC) method~\cite{Maris:2009bx,Maris:2008ax,Bogner:2007rx} which
yields uncertainty estimates for binding energies.

We adopt the traditional harmonic oscillator (HO) basis which enables
us to isolate and remove spurious effects of center of mass (cm)
motion from all observables and from the one-body density matrices (OBDMs)
to high precision.
Since we address light nuclei, we feel this capability is an important
ingredient. A further advantage in using the (HO) basis is its ease in
performing analytical evaluations and straightforward matrix element
calculations though certain alternative basis choices also have these 
advantages~\cite{Keister:1996bd,Caprio:2012}.  In this context, the NCFC
approach is similar to the no-core shell model
(NCSM)~\cite{Navratil:2000gs}.  The main differences are that in the
NCFC approach we do not use the Lee--Suzuki renormalization
procedure~\cite{LSO} which is commonly employed in the NCSM; and more
importantly, we estimate the numerical accuracy of our results based
on the rate of convergence and dependence on the basis space
parameters~\cite{Maris:2008ax,Bogner:2007rx}.

We use a set of finite single-particle HO bases, characterized by two
basis space parameters, the HO energy $\hbar\Omega$ and the many-body
basis space cutoff $N_{\max}$.  $N_{\max}$ is defined as the maximum
number of total oscillator quanta allowed in the many-body basis space
above the minimum for that nucleus.  Independence of both parameters
$\hbar\Omega$ and $N_{\max}$ signals numerical convergence; for bound
states, true convergence is generally expected to be reached in the
limit of a complete (infinite dimensional) basis.  For the binding
energy we use an extrapolation to the complete basis space, with
error bars reflecting the uncertainty due to the extrapolation.

In this work, we look at the actual nucleon densities, in addition to
observables such as spectra, radii, and multipole moments.  In order
to do so, we introduce a technique for unfolding the cm motion from
the one-body density.  This allows us to obtain the
translationally-invariant densities, without any smearing effects from
the cm motion.  Indeed, salient details of the density are often
enhanced in the translationally-invariant (ti) density, compared to the
single-particle densities that are commonly used in configuration
interaction calculations, as we will show.

\section{Methods and Ingredients}

\subsection{Nuclear Hamiltonian}

We begin with the translationally-invariant Hamiltonian for the
$A$-body system in relative coordinates
\begin{eqnarray}
 H_A &=& T_{\hbox{\scriptsize rel}}+V \; = \;  \frac{1}{A}\sum_{i<j}\frac{(\vec{p}_i-\vec{p}_j)^2}{2m} 
\nonumber \\ && {}
 + \sum_{i<j}V_{NN}(\vec{r}_i-\vec{r}_j)
 + \sum_{i<j}V_{C}(\vec{r}_i-\vec{r}_j) \;,
\label{Eq:relHam}
\end{eqnarray}
where $mc^2$ is the nucleon mass times the speed of light (c) squared 
(taken to be 938.92 MeV, the average of the proton and neutron masses), 
$V_{NN}$ is the NN interaction, and $V_C$ is the Coulomb interaction, which acts between
the protons only.  We adopted the NN interaction JISP16, a realistic
NN interaction initially developed from NN data using inverse
scattering techniques.  It is then adjusted with phase-shift
equivalent unitary transformations to describe light nuclei without
explicit three-body
interactions~\cite{Shirokov:2005bk,Shirokov:2004ff,Shirokov:2003kk}.

JISP16 provides good convergence rates for the ground state (gs)
energies of nuclei with $A \leq16$.  We investigate convergence rates
for a selection of additional observables in the present work, in
particular the spectra, radii, and multipole moments.  We use the
naive point-like operators for these observables; we do not take
effects such as meson-exchange currents into account.  However, it is
known that these effects can cause significant corrections to
observables such as magnetic moments~\cite{Marcucci:2008mg}.  We
should therefore expect similar deviations from experiments for our
calculations of these observables.

\subsection{No-Core Full Configuration calculations}

In the many-body framework that we are using, we expand the nuclear
wavefunction $\Psi$ in a basis of Slater determinants of
single-particle HO states.  Note that we use single-particle
coordinates, rather than relative coordinates, in the nuclear
wavefunction.  That means that our wavefunctions, and therefore our
OBDMs calculated as expectation values of one-body
operators, will include cm motion.

All many-body basis states are included with HO quanta up to and
including the amount governed by the $N_{\max}$ truncation.  Thus, if
the highest HO single-particle state for the minimal HO configuration
has $N_0$ HO quanta, then the highest allowed single-particle state in
the truncated basis will have $N_0+N_{\max}$ HO quanta.  Furthermore,
our calculations are 'No-Core' configuration interaction calculations.
This means that all nucleons participate in the interactions on an
equal footing.  As we increase $N_{\max}$, and approach convergence,
we expect physical observables to become independent of both the HO
parameter $\hbar\Omega$ and the truncation parameter $N_{\max}$.
However, due to current limits to our finite basis, our calculations
do show some parameter dependence, even in the largest basis spaces.
We apply previously established extrapolation tools to take the
continuum limit of the binding energy as we discuss shortly.

The HO basis for single-particle states, in combination with this
many-particle $N_{\max}$ truncation, leads to exact factorization of
the nuclear wavefunctions into a cm wavefunction and a
ti wavefunction
\begin{eqnarray}
 \Psi(\vec{r_i}) &=& \Phi^\Omega_{\rm cm}(\vec{R}) \otimes \phi_{\rm ti}
\label{Eq:exactfact}
\end{eqnarray}
where $\vec{R}=(\frac{1}{A})\sum_{i=1}^A\vec{r}_i$ and $\phi_{\rm ti}$
depends only on inter-nucleon coordinates.  The actual Hamiltonian
that we are using involves adding and subtracting $H_{\rm cm}$ so that
$H_A$ takes the form
\begin{eqnarray}
H_A &=& \sum_{i<j}^A\bigg[\frac{(\vec{p}_i-\vec{p}_j)^2}{2mA}+V_{ij}\bigg]
\nonumber \\
    &=& \sum_{i=1}^A\bigg[\frac{p_i^2}{2m}+\frac{1}{2}m\Omega^2r_i^2\bigg]
\nonumber\\
    && {} + \sum_{i<j}^A\bigg[V_{ij}-\frac{m\Omega^2}{2A}(\vec{r}_i-\vec{r}_j)^2\bigg]-H_{\rm cm}
\label{Eq:modHam}
\end{eqnarray}
In order to separate the cm excited states from the low-lying
states of interest, we adopt the Lawson method~\cite{Lawson} whereby
we add a Lagrange multiplier term, 
$\lambda(H^\Omega_{\rm cm} - \frac{3}{2}\hbar\Omega)$, 
to the many-body Hamiltonian, Eq.~(\ref{Eq:modHam})
\begin{eqnarray}
 H&=&H_A+\lambda(H_{\rm cm}-\frac{3}{2}\hbar\Omega) \,.
\end{eqnarray}

With $\lambda$ positive, states with cm excitations are separated by
multiples of $\lambda\hbar\Omega$ from the states with the lowest HO
cm motion.  Since the Lagrange multiplier term acts only on the cm
coordinate, it is independent of the inter-nucleon coordinates and it
does not affect the energy eigenvalues or the
translationally-invariant wavefunctions $\phi_{\rm ti}$ of the
low-lying states.  Indeed, observables for the low-lying states are
independent of $\lambda$, as long as $\lambda\hbar\Omega$ is much
larger than the excitation energy of the highest state of interest.

In the truncated basis space, we can now write the many-body
Schr\"odinger equation as a finite matrix equation with a real,
symmetric, sparse matrix.  The eigenvalues of this matrix give us the
binding energy, and the corresponding eigenvectors give us the
wavefunctions.  In any finite basis space, the eigen-energies satisfy
the variational principle and show uniform and monotonic convergence
from above with increasing $N_{\max}$, allowing for extrapolation to
the infinite basis space.  To obtain the extrapolated gs energy
$E_{gs}(\infty)$, we use a fitting function of the form
 \begin{eqnarray}
  E_{gs}(N_{\max}) &=& a \;\exp(-c\,N_{\max}) + E_{gs}(\infty)\;.
 \end{eqnarray}
This is an empirical method
~\cite{Maris:2009bx,Maris:2008ax,Bogner:2007rx} that is valid within
estimated uncertainties that we now define.  We assign equal weight to
each of three successive values of $N_{\max}$ at a fixed $\hbar\Omega$
and perform a regression analysis.  The difference between
extrapolated results from two consecutive sets of three $N_{\max}$
values is used as the estimate of numerical uncertainty associated
with the extrapolation.  The optimal $\hbar\Omega$ value for this
extrapolation appears to be the $\hbar\Omega$ that minimizes the
difference between the extrapolated energy and the result at the
largest $N_{\max}$.  Typically, this corresponds to a $\hbar\Omega$
value slightly above the variational minimum.  Of course, the
extrapolated results should be independent of $\hbar\Omega$, within
their numerical error estimates, and we do check for such consistency.
Furthermore, we often adjust our numerical error estimate by
considering the results over a range of 5 MeV in $\hbar\Omega$.

For other observables, we do not have a robust and reliable
extrapolation method; we therefore use the degree of (in)dependence
from the basis space parameters $\hbar\Omega$ and $N_{\max}$ as a
measure for convergence as we describe further below on a case-by-case
basis.

For nuclei with mass $A>4$, it is challenging to obtain convergence
(independence of both $N_{\max}$ and $\Omega$) for all observables
within the NCFC approach with realistic interactions.  As $N_{\max}$
increases, the dimension of the many-body basis increases
exponentially, and there is a clear need for high-performance
computing.  For the Li isotopes we investigate here, we have limited
the basis space to about 1 billion: for $^6$Li the largest basis space
is $N_{\max}=16$, with a dimension of 805,583,856; for $^7$Li the
largest basis space is $N_{\max}=14$, with a dimension of
1,244,131,981; and for $^8$Li the largest basis space is
$N_{\max}=12$, with a dimension of 1,222,330,036.  We use the code
MFDn (Many Fermion Dynamics for nuclear
structure)~\cite{Sternberg:2008,Maris:2010iccs}, which is a
state-of-the-art numerical code for no-core configuration interaction
calculations.  The calculations were performed on the Cray XT4 and XE6
at the National Energy Research Supercomputer Center (NERSC), using up
to 8,000 processors (4 or 6 cores/processor) with 8 GB of memory each.

\subsection{One-Body Density Matrix}

The OBDM represents, in a compact form,
sufficient information about quantum states of a system to evaluate
all observables that can be expressed by one-body operators.  For
example, a single wavefunction of $^6$Li at $N_{\max}=16$ has a size
of 5 GB, while the OBDM for a single state in that same basis is less
than 5 MB.  The OBDMs also provide the necessary information for
visualizing nuclear density distributions as we demonstrate below.

The non-local one-body density in coordinate space for an initial
$A$-body wavefunction $\Psi_i$ and final $A$-body wavefunction
$\Psi_f$ is defined as
\begin{eqnarray}
 \rho^{fi}(\vec{r}_1,\vec{r}_1') 
  &=& \int \Psi_f^\star(\vec{r}_1, \vec{r}_2, \ldots, \vec{r}_A) \; 
\nonumber \\ && {} \times
           \Psi_i(\vec{r}_1', \vec{r}_2, \ldots, \vec{r}_A) \;
           d^3r_2 \ldots d^3r_A \,.
\label{Eq:OBDME}
\end{eqnarray}
The limit $\vec{r}_1=\vec{r}'_1$ of Eq.~(\ref{Eq:OBDME}) gives the
local one-body density.  For $\Psi_f = \Psi_i$ this corresponds to the
probability of finding a nucleon at position $\vec{r}_1$ when the
system is in that state.  That is, the local one-body density
distribution is given by
\begin{eqnarray}
 \rho^\Omega_{\rm sf}(\vec{r}_1) 
  &=& \int \rho(\vec{r}_1,\vec{r}_1') \delta(\vec{r}_1 - \vec{r}_1' )  d^3r_1'
\label{Eq:localdens}
\end{eqnarray}
where we suppress the state labels for simplicity and insert a
superscript $\Omega$ to signify the dependence on the HO basis space
used in the evaluation of the eigenfunctions.  Since it depends on the
single-particle coordinates, we refer to this density as the
space-fixed (sf) density.

Note that, due to our use of single-particle coordinates,
rather than relative coordinates, our wavefunctions $\Psi(\vec r_i)$
include cm motion.  The resulting one-body density distributions will
therefore include contributions from the cm motion.  However, because
of the exact factorization of the cm wavefunction and the
ti wavefunction (see Eq.~(\ref{Eq:exactfact}))
this density is actually a convolution of the cm density
$\rho^\Omega_{\rm cm}$ and the ti density
$\rho_{\rm ti}(\vec{r})$
describing the probability of finding a nucleon at position $\vec{r}$
relative to the cm of the entire nucleus\footnote{Note that our ti
  density is not the same as the one-body density in Jacobi
  coordinates, $\sigma(\vec{\xi})$ as defined in Ref.~\cite{giraud:2008},
  which describes the probability of finding one nucleon at $\vec{\xi}$ relative to
  the cm of the remaining $A-1$ nucleons.  Both these densities are
  translationally-invariant, and they are related to each other via
  rescaling factors $\frac{A}{A-1}$.}
\begin{eqnarray}
  \rho^\Omega_{\rm sf}(\vec{r}_1) &=&
     \int \rho_{\rm ti}(\vec{r}_1-\vec{R}) \; \rho^\Omega_{\rm cm}(\vec{R})\; d^3\vec{R}.
\end{eqnarray}
For the HO basis, $\rho^\Omega_{\rm cm}$ is a simple Gaussian (the gs
density of $H_{\rm cm}$) with explicit dependence on $\Omega$ that
smears out $\rho_{\rm ti}$.  This smearing can obfuscate interesting
details of $\rho_{\rm ti}$.  Furthermore, it introduces a spurious
dependence on the basis parameter $\Omega$ into $\rho^\Omega_{\rm sf}$
that masks the convergence.  Even in the limit of a completely
converged calculation, the single-particle density 
$\rho^\Omega_{\rm sf}$ depends on $\Omega$, 
whereas $\rho_{\rm ti}$ becomes independent of the basis.

In order to eliminate these smearing effects and to help develop a
physical intuition for the {\it ab initio} structure of a nucleus, it
would be helpful to see the coordinate space density distributions
free of spurious cm motion.  This can be achieved by a deconvolution
of the cm density and the ti density using
standard Fourier methods~\cite{Morse}
\begin{eqnarray}
 \rho_{\rm ti}(\vec{r}_1)&=& F^{-1}\bigg[
   \frac{F[\rho^\Omega_{\rm sf}(\vec{r}_1)]}{F[\rho^\Omega_{\rm cm}(\vec{R})]}\bigg]
\end{eqnarray}
where $F[f(\vec r)]$ is the 3-dimensional Fourier transform of $f(\vec r)$.
At convergence, the dependence on $\Omega$ should cancel on the RHS of
this equation.  That means that after this deconvolution, we can
better investigate the convergence of the physically-relevant ti density.

In addition to the 3-dimensional ti densities, we also show results for
different multipoles of the ti density, $\rho_{\rm ti}^{(K)}(r)$, defined 
with the same initial and final states, having total angular momentum $J$ 
and magnetic projection $M$, as
\begin{eqnarray}
 \rho_{\rm ti}(\vec{r}) &=&
\sum_{K=0}^{2J} \frac{\langle J M K 0|J M\rangle}{\sqrt{2J+1}}  
  \; Y_{K}^{\star 0}(\hat{r}) \; \rho_{\rm ti}^{(K)}(r). 
\label{Eq:multipoleti}
\end{eqnarray}
These multipoles allow for a better assessment of the numerical
convergence of the densities, as we will illustrate below.  Another
advantage of the multipole expansion is that it allows for a
straightforward determination of the ti density for any 
$M$: the multipoles $\rho_{\rm ti}^{(K)}(r)$ are
independent of $M$.  However, some features of the density,
in particular clustering, are more apparent from the 3-dimensional
plots of $\rho(\vec{r})$ than from plots of the different multipole components.

The results we present here for densities of the Li-isotopes are all
translationally-invariant unless we state otherwise.  
An alternative
method to isolate the ti density has been developed
in~\cite{Navratil:TID}.

\subsection{Observables}

In HO space, the space-fixed OBDM is specified by its matrix elements
\begin{eqnarray}
 \rho_{\beta\alpha}^{fi} &=& 
  \langle \Psi_f | a^\dagger_\alpha a_\beta | \Psi_i \rangle \,,
\end{eqnarray}
where $\alpha$ and $\beta$ stand for a set of single-particle quantum
numbers $(n_\alpha, l_\alpha, j_\alpha, m_\alpha, \tau_{z,\alpha})$
and $(n_\beta, l_\beta, j_\beta, m_\beta, \tau_{z,\beta})$, and we use
the Dirac bra-ket notation to represent the total many-body state
vector.  Once we have obtained the one-body density matrix elements
$\rho_{\beta\alpha}$ (OBDMEs), we can easily calculate observables
that can be expressed as one-body operators.  For initial and final
states with total angular momentum $J_{i,f}$ and possibly additional
quantum numbers $\lambda_{i,f}$, but with the same magnetic projection
$M$, the E2 matrix elements using the canonical one-body
electromagnetic current operator are given by
\begin{eqnarray}
 M_{\rm E2}^{fi} &=& 
 \langle \lambda_f J_f M | {\rm E2} | \lambda_i J_i M \rangle
\nonumber \\
&=& \sum_{\alpha\beta} \; \rho^{fi}_{\beta\alpha} \;
 \langle\alpha| \int r^2 Y_2^0(\hat{r}) d^3r|\beta\rangle \,,
\end{eqnarray}
with $\alpha$ and $\beta$ restricted to the protons only ($\tau_z =
\frac{1}{2}$).  Note that the fact that the OBDM includes cm motion
does not matter for E2 matrix elements (nor for M1 matrix elements
discussed below): the cm wavefunction is a normalized $s$-wave, and does not
contribute to the integral due to the factor $Y_2^0(\hat{r})$.

For comparison with experiments, it is more convenient to convert
these $M$-dependent matrix elements to reduced matrix elements using
the Wigner--Eckart theorem~\cite{Talmi}.  For a proper tensor
operator $T_{Kk}$ the reduced matrix element is defined by
\begin{eqnarray}
\langle\lambda_f J_f||T_K||\lambda_i J_i\rangle &=& 
{\langle\lambda_f J_f M_f|T_{Kk}|\lambda_i J_i M_i\rangle}
\nonumber \\
&& {} \times
 \frac{\sqrt{2J_f + 1}}{\langle J_f M_f Kk|J_i M_i\rangle}
\end{eqnarray}
provided that the Clebsch--Gordan coefficient in the denominator,
$\langle J_fM_fKk|J_iM_i\rangle$ (following the conventions of
Ref.~\cite{Talmi}), is not zero.  In terms of the reduced E2 matrix
elements, reduced E2 transition probabilities are given by~\cite{Bohr}
\begin{eqnarray}
 {\rm B(E2};i \to f)  &=& \frac{1}{2J_i +1}
 \langle \lambda_f J_f || \; {\rm E2} \; || \lambda_i J_i \rangle ^2 \,
\end{eqnarray}
in units e$^2$~fm$^4$.  The quadrupole moment is conventionally
defined through the E2 matrix element for $M = J$
\begin{eqnarray}
 {\cal Q} &=& \bigg(\frac{16\pi}{5}\bigg)^{1/2}
\langle \lambda J M=J | \; {\rm E2} \; | \lambda J M=J \rangle \,
\end{eqnarray}
and can also be expressed in terms of the reduced matrix element
as~\cite{Bohr}
\begin{eqnarray}
 {\cal Q} &=& \bigg(\frac{16\pi}{5}\bigg)^{1/2}\frac{\langle J J 2 0| J J\rangle}{\sqrt{2J+1}}
\langle \lambda J || \; {\rm E2} \; || \lambda J \rangle \,
\end{eqnarray}
in units e~fm$^2$.

The matrix elements for the M1 transitions and magnetic moments
receive contributions both from the proton and neutron intrinsic spins
and from the proton orbital motion.  Again, we consider only the
canonical one-body electromagnetic current operator, in which case
they can be calculated from the OBDMEs
\begin{eqnarray}
\lefteqn{
M_{\rm M1}^{fi} \;=\; \langle \lambda_f J_f M | {\rm M1} | \lambda_i J_i M \rangle}
\nonumber \\ &=&
\sum_{\alpha\beta} \; \rho^{fi}_{\beta\alpha} \;
 \langle\alpha| 
 \frac{1}{2}(1+\tau_z)(L + g_p \sigma) 
 + \frac{1}{2}(1-\tau_z) g_n \sigma |\beta\rangle 
\nonumber
\end{eqnarray}
where $g_p = 5.586$ and $g_n = -3.826$ are the proton and neutron
gyromagnetic ratios in nuclear magneton ($\mu_N$) units; the
quantities $L$, $\sigma$ and $\tau$ represent the conventional orbital
angular momentum, spin and isospin operators.  In terms of the reduced
M1 matrix element, the reduced M1 transition probabilities are given as~\cite{Bohr}
\begin{eqnarray}
 {\rm B(M1};i \to f)  &=& \frac{1}{2J_i +1}
 \langle \lambda_f J_f || \; {\rm M1} \; || \lambda_i J_i \rangle ^2\,
\end{eqnarray}
in units $\mu_N^2$, and the magnetic moment is defined as
\begin{eqnarray}
 \mu &=& \bigg(\frac{4\pi}{3}\bigg)^{1/2}\frac{\langle J J 1 0| J J\rangle}{\sqrt{2J+1}}
 \langle \lambda J || \; {\rm M1} \; || \lambda J \rangle \,
\end{eqnarray}
in units $\mu_N$.

We also present results for the point-proton root-mean-square (RMS) radius, $\langle
r^2_{pp} \rangle^{1/2}$.  This can be calculated either from the
translationally-invariant local density, $\rho_{\rm ti}$, or directly,
as a two-body operator, from the wavefunction in single-particle
coordinates, $\Psi(\vec{r_i})$.  We performed the calculations with
both approaches and confirmed the results were identical to within the
numerical precision of four significant figures.

In order to convert measured nuclear charge radii,
$\langle r^2_{c} \rangle^{1/2}$, to point-proton radii, we use~\cite{Friar:1997}
\begin{eqnarray}
 \langle r_{pp}^2\rangle&=&
\langle r_{c}^2\rangle-R_p^2-\frac{N}{Z}R_n^2-\frac{3\hbar^2}{4M_p^2c^2}
\end{eqnarray}
Here, $R_p^2=0.769(12)~{\rm fm}^2$ is the RMS proton charge radius,
$R_n^2=-0.177(4)~{\rm fm}^2$ the RMS neutron charge radius, $M_p$ the
proton mass, 
and $\frac{3\hbar^2}{4M_p^2c^2}\approx 0.033~{\rm fm}^2$ the
Darwin--Foldy correction.  

In addition to these correction terms there
is also a spin-orbit contribution to the charge radius, but this
contribution is model-dependent, and (for the nuclei discussed here)
less than 1\% for realistic wavefunctions~\cite{Nortershauser:2011zz}.
Hence we neglect it. For the experimental radii we use the values of
Ref.~\cite{Nortershauser:2011zz}, which were obtained from
high-precision laser spectroscopy measurements of isotope shifts in
combination with the $^6$Li charge radius as absolute reference.
Using the $^7$Li charge radius as reference in combination with the
same isotope shifts gives radii that are about 2\% to 3\% smaller than
the ones we have listed in the tables below~\cite{Sanchez:2006}.

\section{Results}

\subsection{$^6$Li}

\begin{figure}[t]
  \includegraphics[width=0.95\columnwidth]{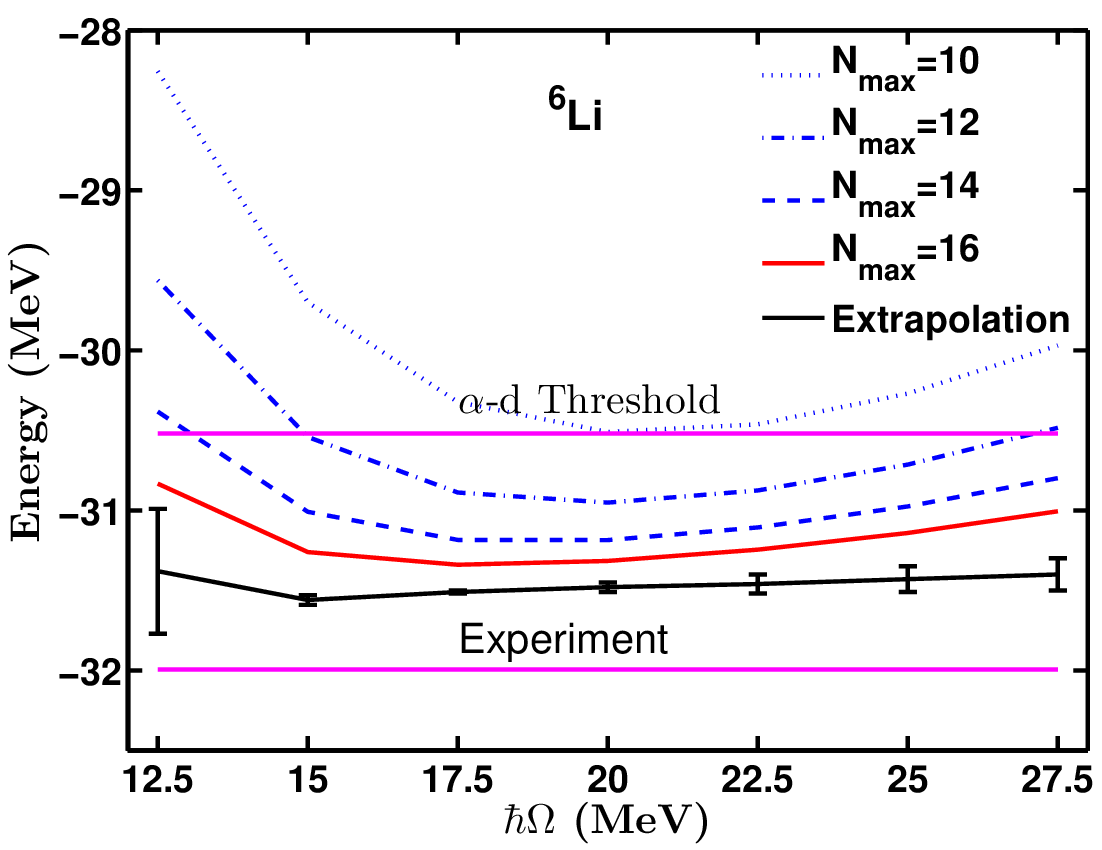}
  \includegraphics[width=0.95\columnwidth]{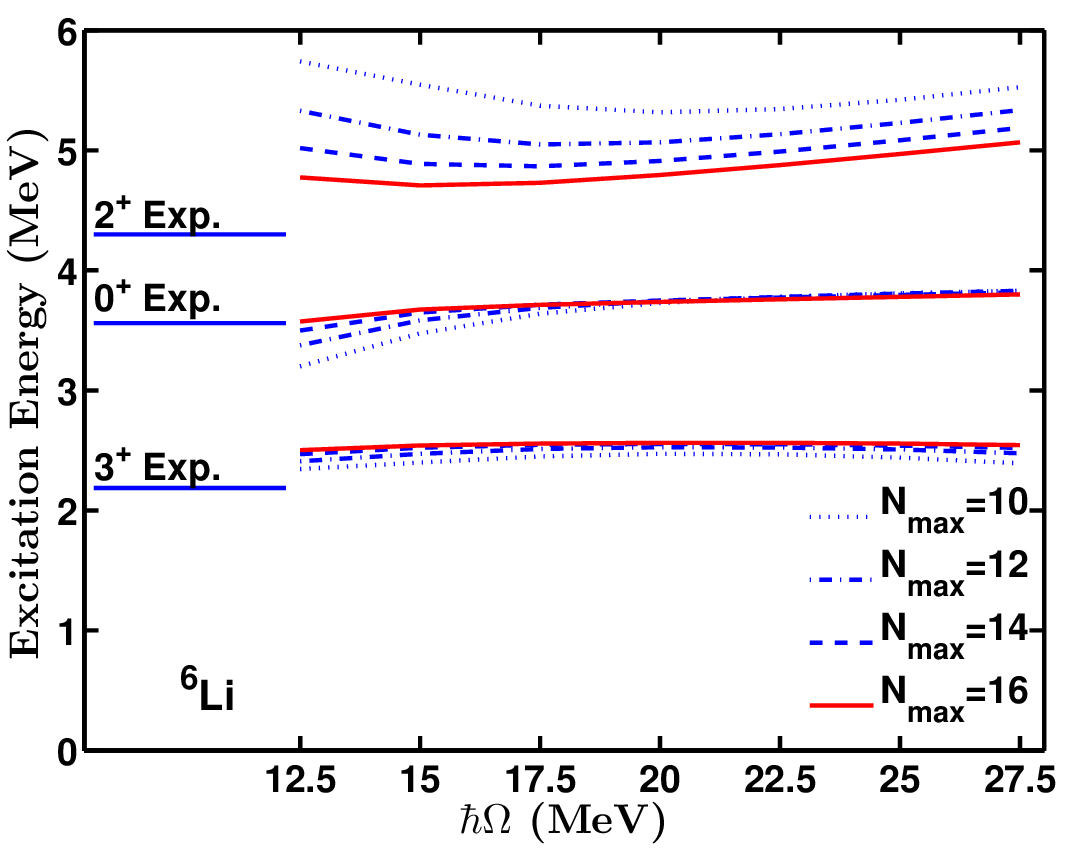}
  \caption{(Color online) 
    The gs energy (top) and excitation spectrum (bottom) of $^6$Li for 
    a sequence of $N_{\max}$ values (indicated in the legend) and as
    function of the HO energy.  The extrapolated gs energy is shown at
    specific values of $\hbar\Omega$ with uncertainties (defined in the
    text) indicated as error bars.
    \label{Fig:energy6Li}}
\end{figure}
In Fig.~\ref{Fig:energy6Li}, we compare the gs energy and excitation
energies at a sequence of $N_{\max}$ values and as a function of the
HO energy $\hbar\Omega$.  We also provide the extrapolated gs energy
as a function of $\hbar\Omega$ along with the assessed uncertainties
(error bars) as described above.

The gs energy for $^6$Li is rapidly converging as indicated by the
emerging independence of the two basis parameters ($N_{\max}$,
$\hbar\Omega$).  The convergence is most rapid around
$\hbar\Omega=17.5$ to $20$~MeV, where the variational upper bound on
the energy is minimal.  Our extrapolated gs energy~\cite{Maris:2008ax}
shows that the system is underbound by 0.50~MeV.  Excitation energies
are well converged at higher $N_{\max}$ (12 and above) values, at
least for $3^+$ and $0^+$ states.  Note that these states are narrow
resonances: the experimental width of the $3^+$ is 24 keV and the
width of the first excited $0^+$ is 8 eV.

The excitation energy of the first $2^+$ state is much less converged,
and shows a systematic increase with increasing $\hbar\Omega$.  Such
$\hbar\Omega$-dependence of the excitation energy is typical for wide
resonances as observed in comparisons of NCSM results with inverse
scattering analysis of $\alpha$-nucleon scattering
states~\cite{Shirokov:2008jv,Maris:2009bx}.  In light of these
previous analyses, the significant $\hbar\Omega$-dependence seems
commensurate with the large experimental width of $1.3$~MeV for this
$2^+$ state.

\begin{figure}[t]
\includegraphics[width=0.95\columnwidth]{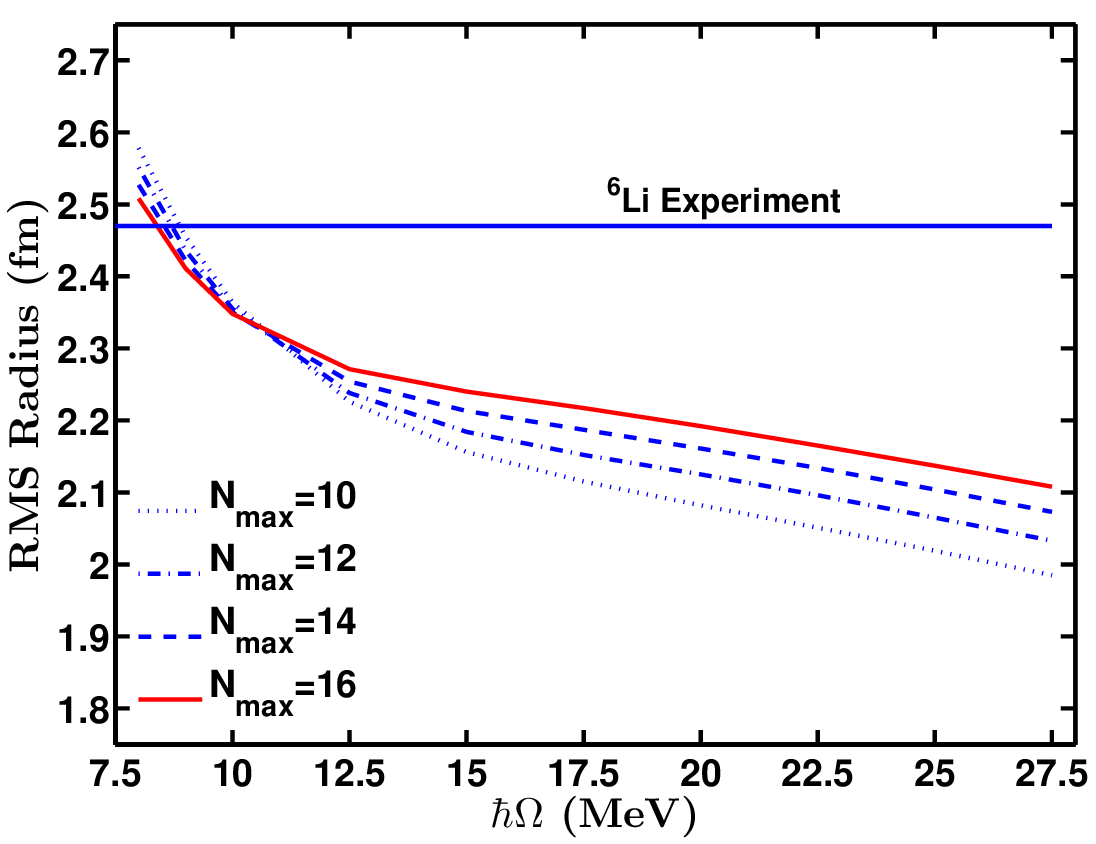}
\includegraphics[width=0.95\columnwidth]{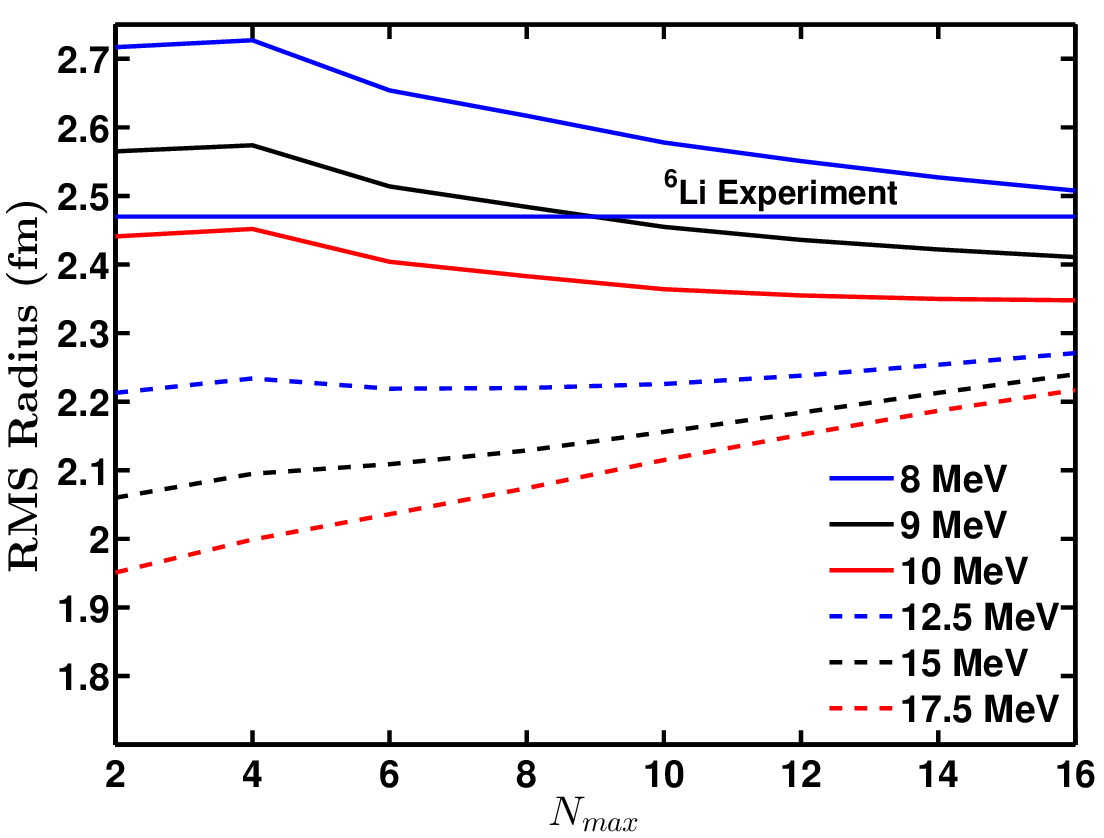}
\caption{(Color online) 
  The RMS point-proton radius of the gs of $^6$Li as a function of HO
  energy at various $N_{\max}$ values (top) and as function of
  $N_{\max}$ at various values of the HO energy (bottom).
  \label{Fig:rad6Li}}
\end{figure}

In Fig.~\ref{Fig:rad6Li}, we show the dependence of the RMS
point-proton radius on the basis space parameters $N_{\max}$ and
$\hbar\Omega$ for $^6$Li.  It appears that this radius is converging
less rapidly than the gs and excitation energies.  Furthermore, the
convergence is neither monotonic nor uniform: at small values of the
HO energy the radius tends to decrease with increasing $N_{\max}$,
whereas at larger values of $\hbar\Omega$ the radius increases with
increasing $N_{\max}$; around $\hbar\Omega=10$ to $12.5$~MeV the RMS
radius is nearly independent of $N_{\max}$.  Because of this, it is
difficult to make realistic estimates of uncertainties for radii and
other long-range observables such as quadrupole moments which exhibit
similar patterns.

The convergence patterns shown in Fig.~\ref{Fig:rad6Li} may be
understood from the following observation: since the HO wavefunctions
fall off like a Gaussian, ${\rm e}^{-c r^2}$ while the true nuclear
wavefunction falls off like an exponential, ${\rm e}^{-d r}$,
observables whose calculations are weighted towards the tail of the
wavefunction, such as the RMS radius, will converge slower than those
observables that depend less on the tails, such as the energy and the
magnetic moment (see below).  Furthermore, it is well known that the
RMS radius, and also other long-range operators such as the quadrupole
moment, are minimally affected by the short-range
correlations~\cite{Stetcu:2006zn,Roth:2010bm}.  And the value of
$\hbar\Omega$ that minimizes the gs energy is not necessarily the
value of $\hbar\Omega$ that best represents the long-range behavior of
the wavefunction.

\begin{figure}[t]
\includegraphics[width=0.95\columnwidth]{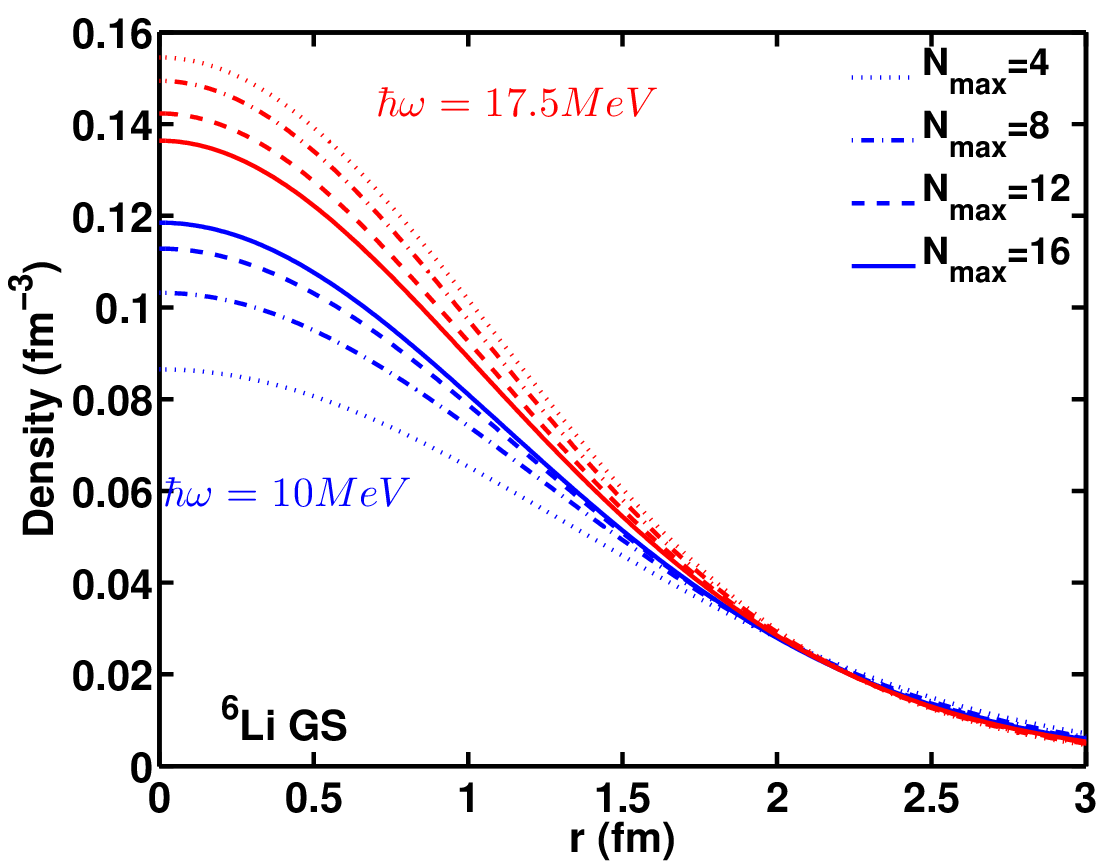}
\includegraphics[width=0.95\columnwidth]{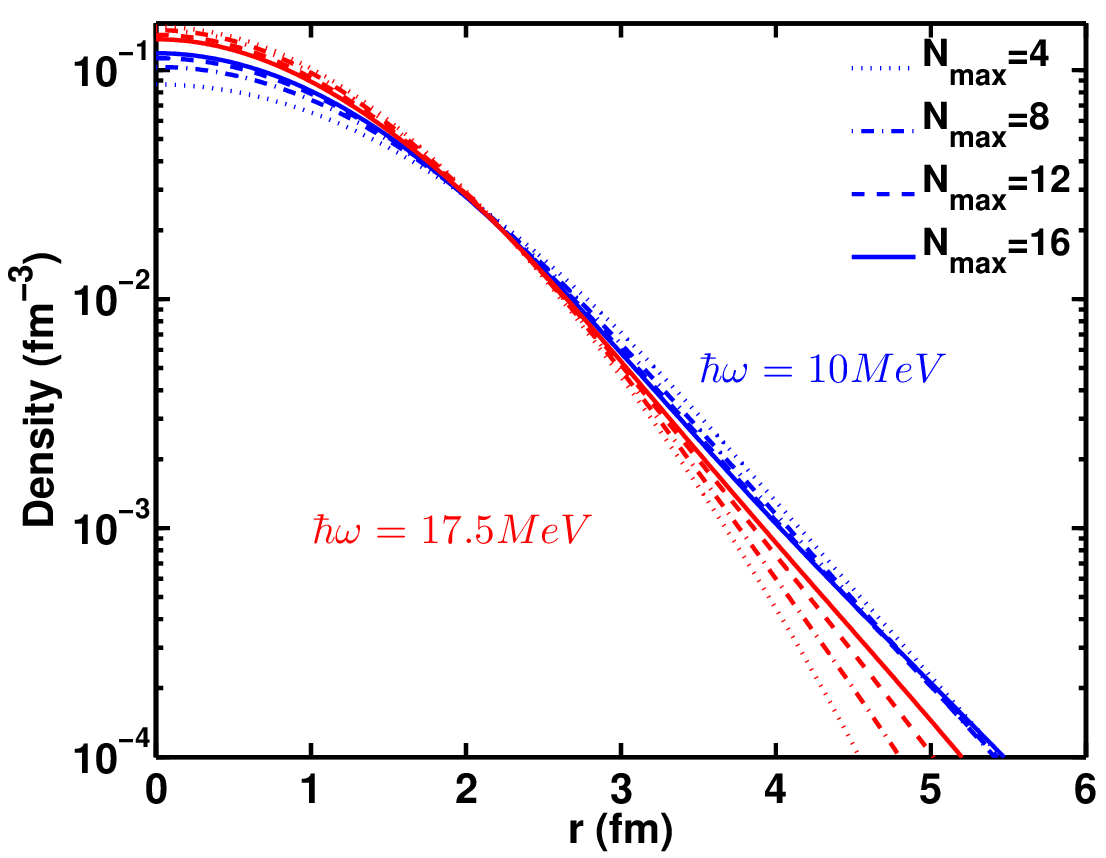}
\caption{(Color online)
  The angle-averaged density
  of the $^6$Li gs for various $N_{\max}$
  values at $\hbar\Omega=10$ MeV (blue curves) and $17.5$ MeV 
  (red curves) on on a
  linear (top) and semi-logarithmic (bottom) scale.  
  \label{Fig:convdens6Li}}
\end{figure}

In Fig.~\ref{Fig:convdens6Li} we show the radial density distribution
for two sets of finite basis spaces, with $\hbar\Omega=10$ and $17.5
$~MeV respectively.  The lower panel shows that the exponential tail
is much better represented in a HO basis with $\hbar\Omega=10$ MeV than in
a HO basis with $17.5 $~MeV, and that the long-range behavior of the
one-body density is therefore converging much more rapidly in a HO
basis with $\hbar\Omega=10$ MeV, and rather poorly converging in the HO
basis that minimizes the gs energy.  That is, the radial density
calculated with $\hbar\Omega=10$ MeV shows much more consistent
long-range behavior at the three highest $N_{\max}$ values than the
density calculated with $\hbar\Omega=17.5$ MeV.  This leads us to the
conclusion that while the value of $\hbar\Omega$ that minimizes the gs
energy is an appropriate value when calculating gs and excitation
energies, as well as magnetic observables, this value is not necessarily optimal 
for calculations of observables that depend primarily on
long-range correlations, even in the moderately large basis spaces
considered here.

Therefore we will quote results for long-range observables at the
$\hbar\Omega$ value where the RMS radii for various $N_{\max}$ values
intersect as seen in the top portion of Fig.~\ref{Fig:rad6Li}, rather
than at the $\hbar\Omega$ value that minimizes the gs energy.  To be
specific, for the Li-isotopes under discussion here, we simply take
the results at $\hbar\Omega=10$ to $12.5$ MeV (where the $N_{\max}$
dependence appears to be minimal) as our approximation to the
converged value of the RMS radius.  In a similar fashion, we will cite
results for the region of minimal $N_{\max}$ dependence for other
observables that depend primarily on long-range correlations.  Such
observables include RMS radii, E2 moments, and B(E2) transitions.
Robust extrapolations to the infinite basis space and reliable error
estimates for these observables remain an open question.

\begin{figure}[t]
\includegraphics[width=0.95\columnwidth]{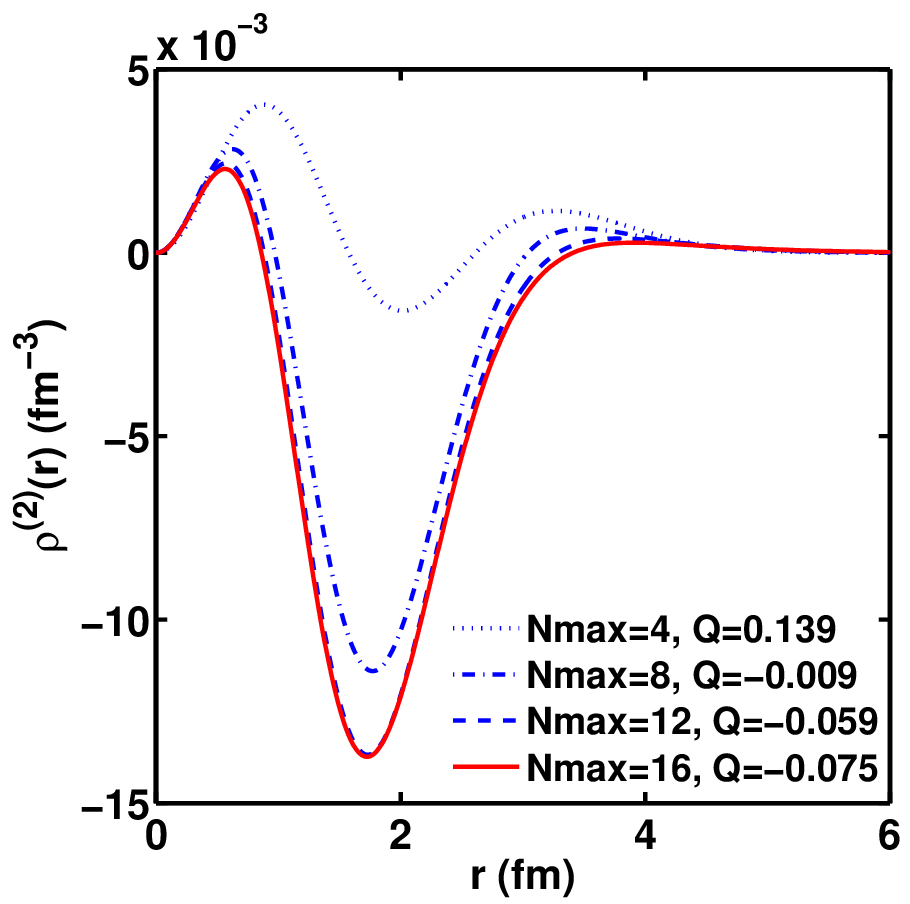}
\includegraphics[width=0.95\columnwidth]{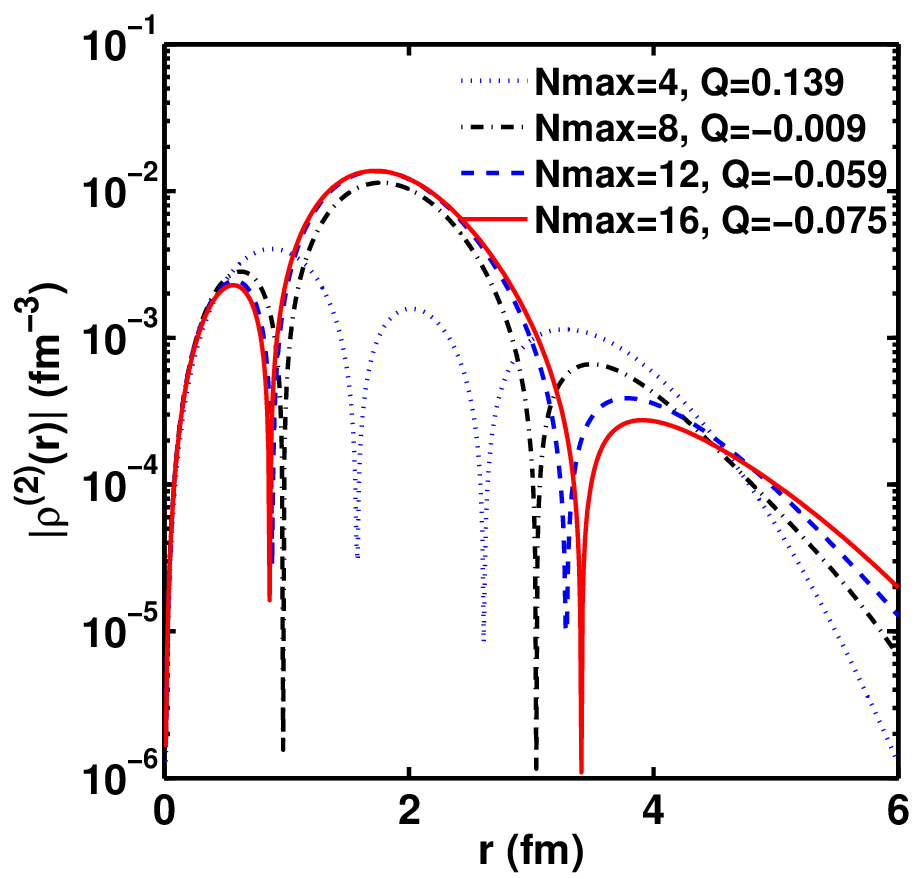}
\caption{(Color online)
  The radial quadrupole density $\rho^{(2)}_{\rm ti}(r)$
  of the $^6$Li gs for various
  $N_{\max}$ values at $\hbar\Omega=17.5$ MeV
on a linear (top) and semi-logarithmic (bottom) scale.
\label{Fig:quad6Li}}
\end{figure}

In Fig.~\ref{Fig:quad6Li} we show the radial quadrupole component of
the $^6$Li gs density distribution, $\rho^{(2)}_{\rm ti}(r)$, at
$\hbar\Omega=17.5$~MeV for several $N_{\max}$ values.  From this figure
it is evident that the quadrupole moment
\begin{eqnarray}
  {\cal Q} &=& \frac{4}{5}\sqrt{\frac{\pi}{6}} \; \int \rho^{(2)}(r) \; r^4dr
\end{eqnarray}
receives both positive and negative contributions.  For $N_{\max}=4$,
the positive contributions are significantly larger than the small
negative contribution around 2~fm, but as $N_{\max}$ increases, this
negative region becomes more pronounced.  At $N_{\max} = 16$ the
quadrupole density is nearly converged inside about 3 fm. The
corresponding quadrupole moment, ${\cal Q} = -0.075$~e~fm$^2$, 
appears to approach convergence.  The small positive region inside 0.9~fm
contributes less that 1\% to this value, but about half of the
contribution from the negative region between 0.9~fm and 3.4~fm is
cancelled by the positive tail of the quadrupole distribution: their
contributions are $-0.167$~e~fm$^2$ and $+0.092$~e~fm$^2$
respectively.  Note that this positive tail is well outside the charge
radius.

\begin{table}[t]
\renewcommand{\arraystretch}{1.2}
\begin{ruledtabular}
\begin{tabular}{lcrrrr}
  $^6$Li              & Expt.     & JISP16     &  AV18/IL2 & CD-B  & INOY \\ \hline
  $E_{\rm b}(1^+,0)$    & 31.994    & 31.49(3)   &  32.0(1)  & 29.07 & 32.3(2)\\
  $\langle r^2_{pp}\rangle^{1/2}$ & 2.45(5)& 2.3  &   2.39(1) & 2.25  & 2.14 \\[5pt]
  $E_{\rm x}(3^+,0)$    & 2.186(2)   &  2.56(2)    &  2.2(2)    \\ 
  $E_{\rm x}(0^+,1)$    &  3.56(1)   &  3.68(6)    &  3.4(2)    \\
  $E_{\rm x}(2^+,0)$    & 4.312(22)  &  4.5(3)     &  4.2(2)    \\
  $E_{\rm x}(2^+,1)$    & 5.366(15)  &  5.9(2)     &  5.5(2)    \\[5pt]
  $Q(1^+,0)$   & -0.082(2) & -0.077(5)   & -0.32(6) & -0.066 & 0.080\\
  $Q(3^+,0)$   & -         & -4.9        &        \\[5pt]
  $\mu(1^+,0)$ & 0.822   & 0.839(2)    & 0.800(1) & 0.843 & 0.843\\
  $\mu(3^+,0)$ & -       & 1.866(2)    &          \\[5pt]
  B(E2;$(3^+,0)$) & 10.7(8)  & 6.1 & 11.65(13) \\
  B(E2;$(2^+,0)$) & 4.4(23)  & 7.5 &  8.66(47) \\[5pt]
  B(M1;$(0^+,1)$) & 15.43(32)& 14.2(1)  & 15.02(11) \\
  B(M1;$(2^+,0)$) &  -       &$<$ 0.001 & 0.002(1)  \\
  B(M1;$(2^+,1)$) & 0.1 (3)  & 0.05(1)  &           \\[5pt]
  $M_{\rm GT}$                          & 2.170    & 2.227(2) & 2.18(3)  \\
\end{tabular}
\end{ruledtabular}
\caption{
  Selected $^6$Li observables calculated up through $N_{\max}=16$. 
  The energies are in MeV; the RMS point-proton radius is in fm; 
  the quadrupole moments are in e~fm$^2$;
  the magnetic moments are in $\mu_N$;
  the reduced B(E2) transition probabilities are in ${\rm e}^2{\rm fm}^4$; 
  and the reduced B(M1) transition probabilities are in $\mu_N^2$.
  All listed transitions are to the ground state.
  The energies are obtained from extrapolations to the infinite basis
  space, with error estimates as discussed in the text;
  the dipole observables as well as the gs quadrupole moment are
  converged within the quoted uncertainty;
  the other quadrupole observables and the RMS point-proton 
  radius are evaluated at $\hbar\Omega=12.5$~MeV.
  We used Ref.~\cite{Nortershauser:2011zz} for the experimental value
  of the RMS radius and Ref.~\cite{Vaintraub} for GT matrix element;
  the other experimental values are from Refs.~\cite{F19881,Tilley20023}.
  AV18/IL2 results are from Refs.~\cite{Marcucci:2008mg,Pervin:2007,Pieper:2004qw,Pieper:2001ap}
  and include meson-exchange corrections for the dipole observables;
  CD-Bonn and INOY results are from Ref.~\cite{CDBINOY} 
  calculated at $N_{max}$=16 and $\hbar\Omega$=11 and 14 MeV
  respectively, with the INOY gs energy
  extrapolated to the infinite basis space.
  \label{Tab:6Li}}
\end{table}

Table~\ref{Tab:6Li} presents a capsule view of selected spectral and
other observables for $^6$Li.  In principle, we can extrapolate not
only the gs energy, but also the binding energies of the excitated
states to the infinite basis space.  We can then calculate the
excitation energy as the difference between the extrapolated binding
energies, treating the numerical error estimates as independent.  For
the excitation energies that appear to be converged (in particular the
$3^+$ state in Fig.~\ref{Fig:energy6Li}), such a procedure leads to an
overestimate of the numerical uncertainty in the excitation energy:
part of the numerical error is common to the ground state and the
excited state.  For such states we reduce the obtained numerical error
estimates based on the apparent convergence of the excitation energies
in finite basis spaces.  On the other hand, for states with excitation
energies that show a significant $\hbar\Omega$ dependence (such as the
$2^+$ state in Fig.~\ref{Fig:energy6Li}), we increase our numerical
error estimate based on this $\hbar\Omega$ dependence.

In general, magnetic dipole observables tend to converge rapidly.
Indeed the magnetic moments of the gs and first excited state, as well
as the B(M1) transitions to the gs, are well converged, with, at
$N_{\max}=16$, a residual dependence on the basis space parameters
that is of the same order as the overall accuracy of our 
lowest eigenvalue.

For the RMS point-proton radius, as well as the B(E2) transitions to
the gs, we list our results from the largest basis space,
$N_{\max}=16$, at $\hbar\Omega=12.5$~MeV, as discussed above.  The
electric quadrupole moment of the gs is 
in excellent agreement with the experimental value - perhaps
better than might be expected due to basis space limitations and
long-range nature of the quadrupole moment operator.

Both the gs binding energy and the excitation energies calculated with
JISP16 compare favorably to those calculated with alternative
realistic NN interactions, i.e. Argonne V18, 
CD-Bonn, INOY,
and SRG evolved N3LO
interactions~\cite{Pieper:2004qw,CDBINOY,Navratil:2001zz,Roth:2010bm}.
However, with the addition of appropriate three-body interactions,
such as Illinois-2~\cite{Pieper:2004qw,Pieper:2001ap} or chiral
three-body forces~\cite{vanKolck:1994yi,Epelbaum:2002vt,Entem:2003ft},
one can obtain somewhat better agreement with data than with the
NN-only interaction JISP16.  

The obtained RMS point-proton radius is
similar to that obtained with CD-Bonn: both are about 10\% too small
compared to experiment.  INOY~\cite{CDBINOY} gives an even smaller
radius, whereas AV18 plus Illinois-2 leads to a
radius~\cite{Pieper:2001ap} that is closer to the experimental
value.  Note however that the gs quadrupole moment obtained with that
interaction is significantly larger than experiment~\cite{Pieper:2001ap}.

Also the calculated magnetic moment and reduced M1 transition probabilities
with JISP16 compare well with available experimental data for all the cases
shown in Table~\ref{Tab:6Li}.

The $(0^+, 1)$ excited state of $^6$Li is the isobaric analog of the
ground state of $^6$He, and can be used to calculate the Gamow--Teller
transition between $^6$Li and $^6$He.  Assuming isospin symmetry, the
Gamow--Teller matrix element $M_{\rm GT}$ is to good approximation
given by
\begin{eqnarray}
M_{\rm GT} &=& \sum_{\alpha,\beta}  \; \rho_{\beta\alpha}^{fi} \;
\langle\alpha|\sigma\tau_+|\beta\rangle  
\end{eqnarray}
and is related to the half-life~\cite{Vaintraub} through
\begin{eqnarray}
 |M_{\rm GT}|^2 &=& 
 \frac{1}{\frac{f_A}{f_V}g_A^2}\frac{2\pi^3\ln2/(G^2|V_{ud}|^2)}{(fT_{1/2})_tm_e^5}
\end{eqnarray}
where $g_A=1.2695(29)$ is the axial constant,
$\frac{f_A}{f_V}=1.00529$ accounts for the difference in the
statistical rate function of the vector and axial-vector transitions,
$m_e$ is the mass of the electron, $G=1.166371(6)\cdot10^{-11} {\rm
  MeV}^{-2}$ is the Fermi coupling constant, and $V_{ud}=0.9738(4)$ is
the CKM matrix element that mixes the quarks involved in the decay.
Our $M_{\rm GT}$ result, presented in Table~\ref{Tab:6Li}, compares
quite well to that calculated in~\cite{Vaintraub} using the
hyperspherical-harmonic expansion method with the same (JISP16)
interaction; they also obtained a value of $M_{\rm GT}=2.227$.  It is
interesting to note that Ref.~\cite{Vaintraub} found the exchange
current corrections to the GT matrix element to be of the order of a
few percent.  Ref.~\cite{Vaintraub} also presents 
the extrapolated gs binding energy of $^6$Li with JISP16 as 31.46(5) MeV 
which is in excellent agreement of our 31.49(3) MeV result presented 
in Table ~\ref{Tab:6Li}. 

\subsection{Density distributions}

A closer look at the 3-dimensional one-body densities, free of
spurious cm effects, would help to develop a physical intuition for
the {\it ab initio} structure of a nucleus.  Although it is easier to
perform the deconvolution of the cm density after integrating out all
angle-dependence one can also deconvolute the full 3-dimensional
density (see the appendix for details).  However, a detailed
investigation of the numerical convergence is impractical for these
3-dimensional densities. We therefore present all our 3-dimensional
density distributions in the largest basis space at
$\hbar\Omega=12.5$~MeV only, where the RMS radius, as well as the
quadrupole moments (which are closely related to the shape of the
wavefunction), appear to be reasonably converged.

\begin{figure}[t]
\includegraphics[width=0.95\columnwidth]{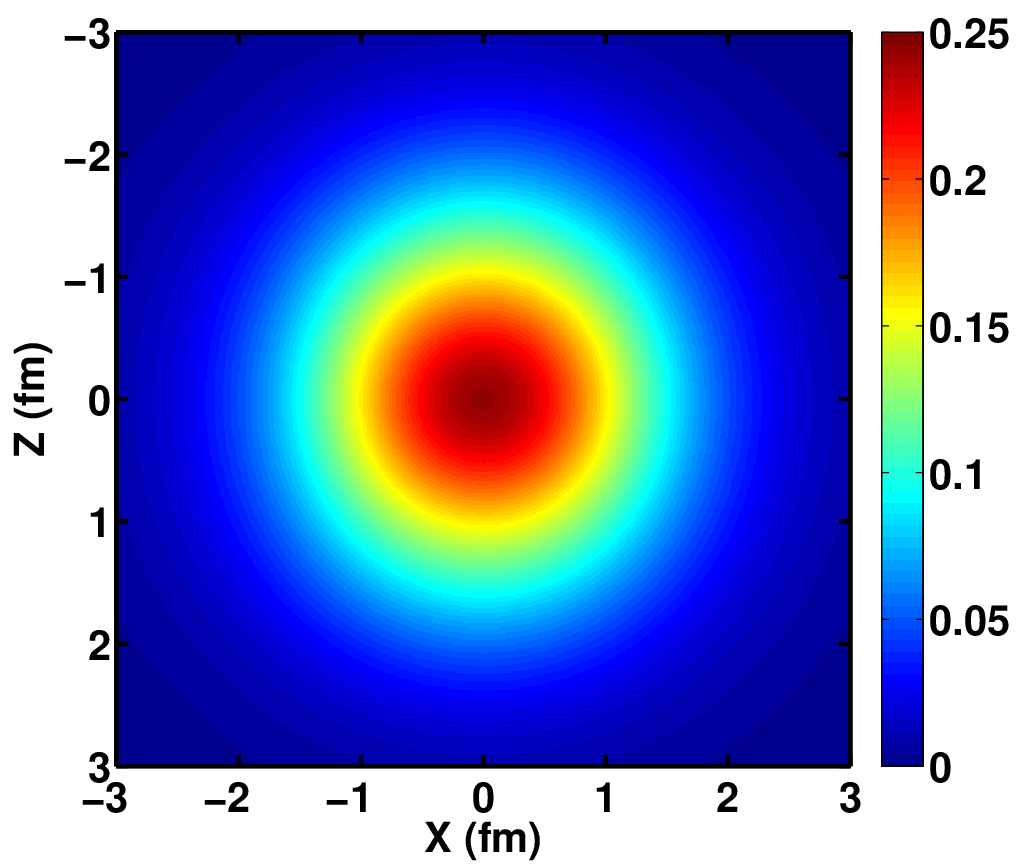}
\includegraphics[width=0.95\columnwidth]{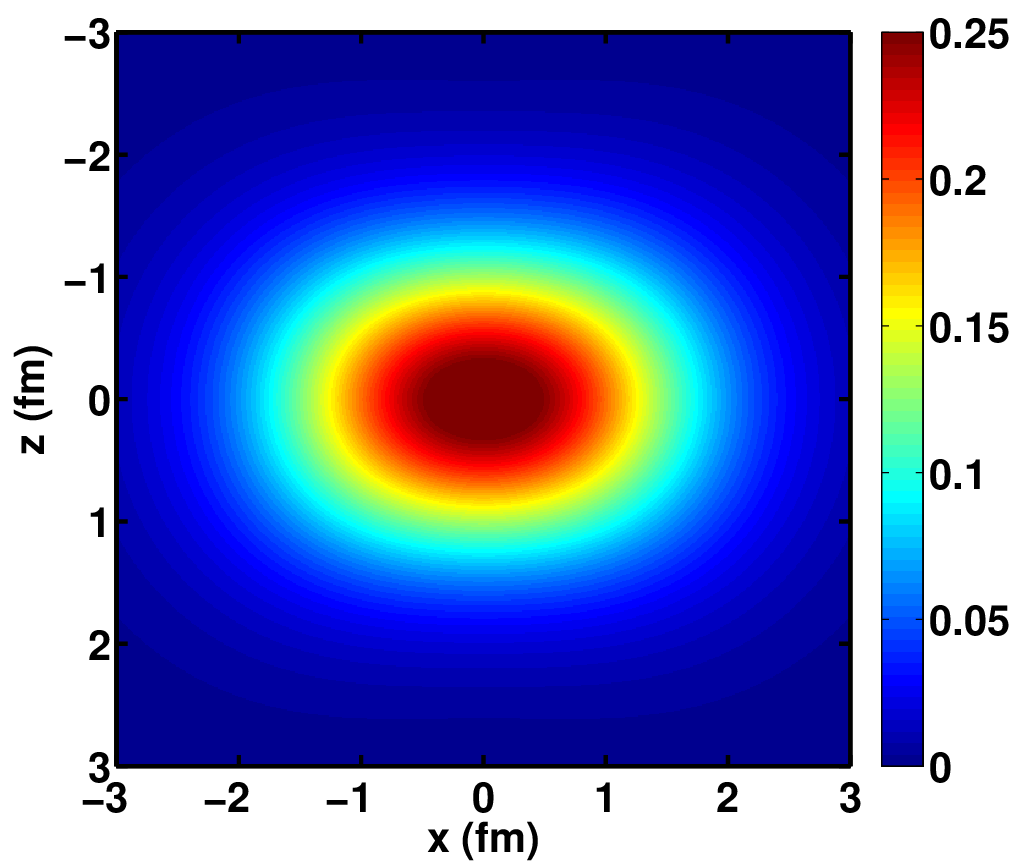}
\caption{(Color online)
  The $y=0$ slice of the translationally-invariant matter density in
  the $x$-$z$ plane for the gs of $^6$Li (top, $J=1$) is contrasted
  with the density for the first excited state (bottom, $J=3$).  
  These densities were calculated at $N_{\max}=16$ and
  $\hbar\Omega=12.5$~MeV.
  \label{Fig:dens3D6Li}}
\end{figure} 

In order to produce the density that represents the actual shape of a
specific state of a nucleus in a translationally-invariant (inertial)
frame, we set $M_J=J$ for all our calculations of the local density.
That is, we select the maximal positive angular momentum projection along the axis
of quantization, the $z$-axis.  This seems like the natural choice
since the quadrupole moment $Q$ is defined as the E2 matrix element at
$J=M_J$ (or equivalently, to the reduced E2 matrix element).  Note
that even though we calculate 3-dimensional density distributions, our
results are symmetric under rotations around the $z$-axis: the
wavefunctions have azimuthal symmetry.

Fig.~\ref{Fig:dens3D6Li} shows the matter density for the lowest two
states of $^6$Li.  Both states are oblate, though the $(1^+,0)$ state
is nearly spherical whereas the $(3^+,0)$ state is strongly oblate.
Indeed, the relative deformation of the translationally-invariant
densities for the gs and first excited state is implied by the results
in Table~\ref{Tab:6Li} for their respective quadrupole moments. The gs
has a negative calculated quadrupole that is near zero, in close
agreement with experiment.  On the other hand, the first excited state
has a large negative calculated quadrupole moment.

\begin{figure}[tp!]
\includegraphics[width=0.75\columnwidth]{Li6_Nm16_MJ3_hw12_5_3Plus_TIDens.eps}
\includegraphics[width=0.75\columnwidth]{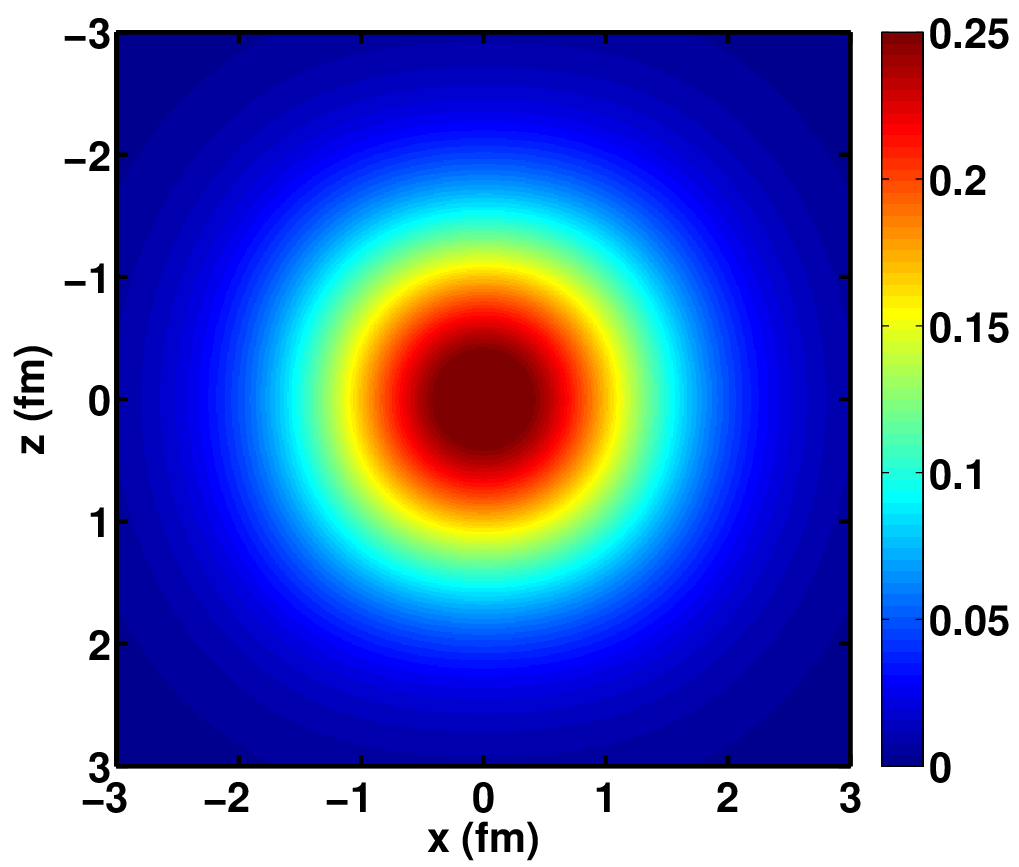}
\includegraphics[width=0.75\columnwidth]{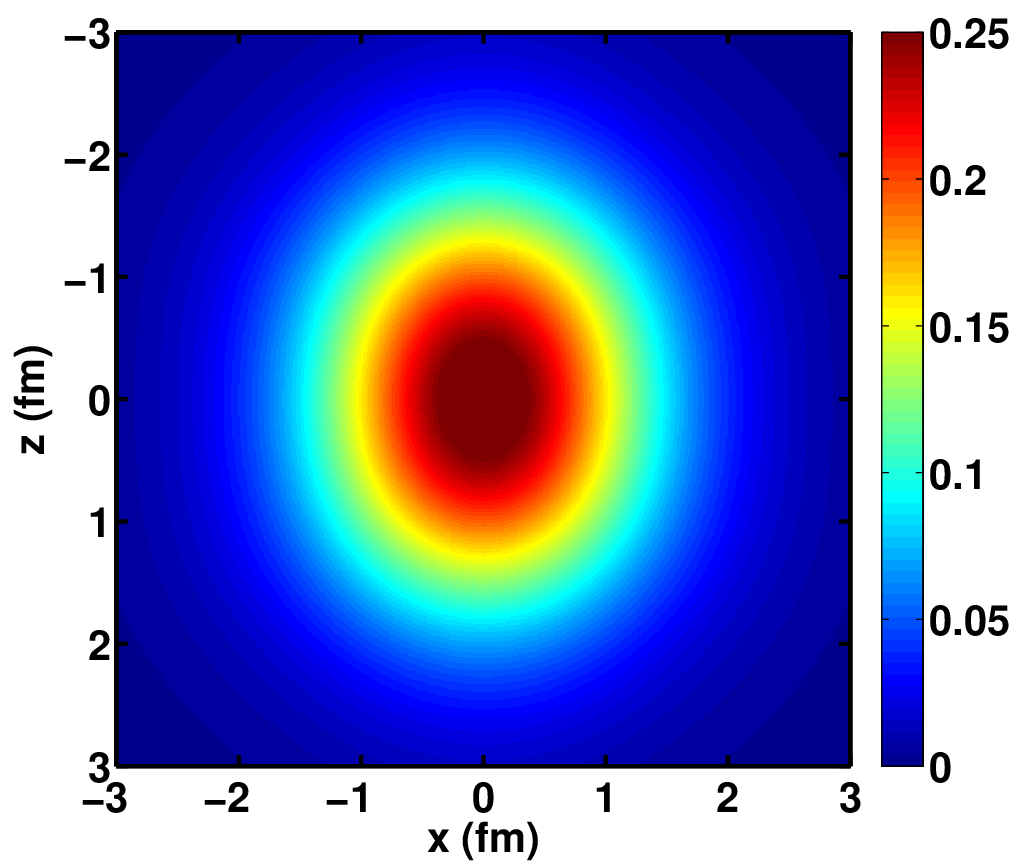}
\includegraphics[width=0.75\columnwidth]{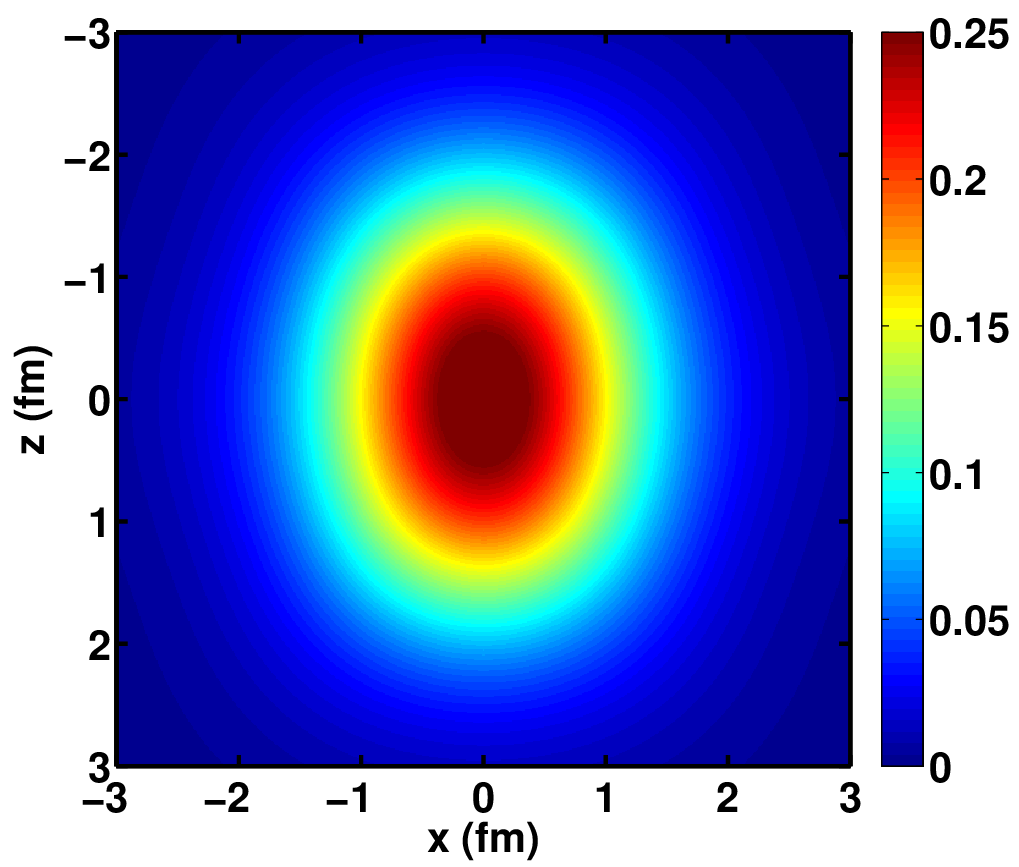}
\caption{(Color online)
  The $y=0$ slice of the translationally-invariant matter density in
  the $x$-$z$ plane for first excited $3^+$ state of $^6$Li with
  $M_j=3$, $2$, $1$, $0$ from top to bottom.  
  \label{Fig:dens3D6LiMj}}
\end{figure} 

It is worth commenting that our use of the terms ``prolate" and
``oblate" characterize the shapes in the inertial frame, not a
body-fixed axis as is common for discussions of shapes in the
collective model~\cite{Bohr}.  In the inertial frame of reference
positive quadrupole moments correspond to prolate shapes and negative
quadrupole moments correspond to oblate shapes.

Fig.~\ref{Fig:dens3D6LiMj} illustrates the effect of $M_J$ on the
density distribution as we see the oblate shape of the density at
$M_J=3$ (top panel) morph into the prolate shape at $M_J=0$ (bottom
panel).  Calculating the density when $J\neq M_J$ gives a density
whose azimuthal symmetry axis is not aligned with the spin.  E.g. at
$M_J=0$ the symmetry axis lies in  the $x$-$y$ plane, perpendicular to
the $z$-axis.  The oblate shape we found at $M_J=3$ is now also
perpendicular to the $z$-axis.  On the other hand, we do have an
azimuthal symmetry around the $z$-axis.  Therefore, what we obtain is
an oblate shape perpendicular to the $z$-axis, but with its principle
axis (symmetry axis) averaged over all directions in the $x$-$y$
plane.  This results in a prolate shape at $M_J=0$, as we see in the
bottom panel of Fig.~\ref{Fig:dens3D6LiMj}.  Note that we see the same
for e.g. the E2 matrix element: with $M_J=0$, the E2 matrix element
for this state is positive, but the corresponding quadrupole moment,
shown in Table~\ref{Tab:6Li}, is negative (and independent of $M_J$).

\subsection{$^7$Li}

\begin{figure}[t]
\includegraphics[width=0.95\columnwidth]{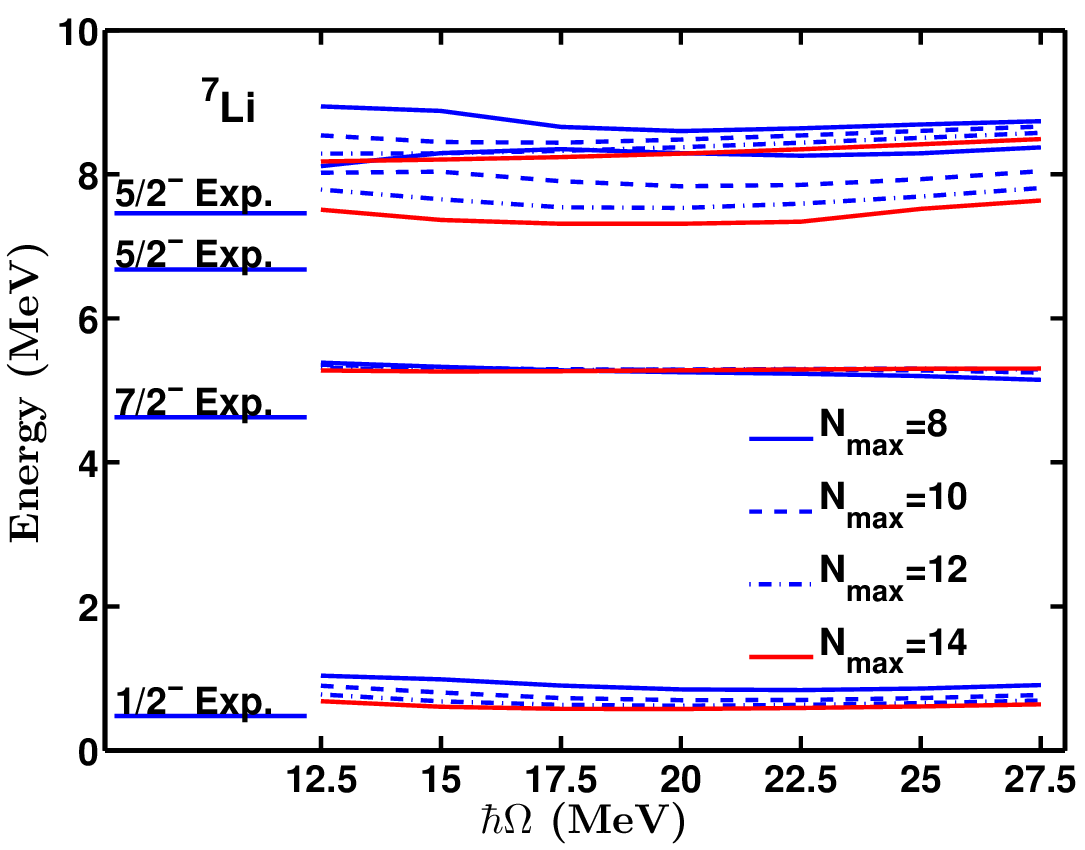}
\caption{(Color online)
  Excitation energies for selected excited states of $^7$Li are
  shown as a function of $\hbar\Omega$ at various $N_{\max}$ values as
  indicated in the legend, and compared to experiment.
  \label{Fig:energy7Li}}
\end{figure}

For $^7$Li we evaluated the low-lying states in basis spaces up to
$N_{\max}=14$.  We only consider isospin $\frac{1}{2}$ states -- the
lowest isospin $\frac{3}{2}$ has more than $10$ MeV excitation energy.
The lowest five states in the excitation energy spectrum compare well
with experiment and the correct level ordering is preserved as shown
in Fig.~\ref{Fig:energy7Li}.  The excitation energy of four of these
five states shows rapid numerical convergence with $N_{\max}$ and
stability with respect to variations in the HO energy.  However, the
convergence of the lower of the two $\frac{5}{2}^-$ states is
significantly slower.  Indeed, experimentally this state has a large
width of $0.88$~MeV, whereas the width of the other states is less
than $0.1$~MeV.  Thus, as in $^6$Li, we again observe a good
correlation between experimental width and convergence rate of
excitation energies.

\begin{table}[t]
\renewcommand{\arraystretch}{1.2} 
\begin{ruledtabular}
\begin{tabular}{lcrrrr}
  $^7$Li                   & Expt.  & JISP16   &  AV18/IL2 & CD-B  & INOY \\ \hline
  $E_{\rm b}(\frac{3}{2}^-)$ & 39.244 & 38.57(4) & 38.9(1)   & 35.56 & 39.6(4) \\
  $\langle r^2_{pp}\rangle^{1/2}$ & 2.30(5) & 2.2 & 2.25(1)   & 2.22  & 2.05 \\[5pt]
  $E_{\rm x}(\frac{1}{2}^-)$  &  0.477   & 0.52(6) &  0.2(1) \\
  $E_{\rm x}(\frac{7}{2}^-)$  & 4.630(1) & 5.25(5) &  4.9(1) \\ 
  $E_{\rm x}(\frac{5}{2}^-_1)$ & 6.680(50) & 7.1(2) &  6.6(1) \\ 
  $E_{\rm x}(\frac{5}{2}^-_2)$ & 7.460(10) & 8.1(1) &  7.2(1) \\[5pt] 
  $Q(\frac{3}{2}^-)$     & -4.06(8)    & -3.2  & -3.6(1) & -3.20 & -2.79\\
  $Q(\frac{7}{2}^-)$     & -           & -5.0  &     \\
  $Q(\frac{5}{2}^-_1)$   & -           & -6.0  &     \\
  $Q(\frac{5}{2}^-_2)$   & -           &  2.3  &     \\[5pt]
  $\mu(\frac{3}{2}^-)$   & 3.256    &  2.954(5) & 3.168(13) & 3.01 & 3.02\\
  $\mu(\frac{1}{2}^-)$   & -        & -0.76(1)  & \\
  $\mu(\frac{7}{2}^-)$   & -        &  3.3(1)   & \\
  $\mu(\frac{5}{2}^-_1)$ & -        & -0.90(2)  & \\
  $\mu(\frac{5}{2}^-_2)$ & -        & -0.39(5)  & \\[5pt]
  B(E2;$\frac{1}{2}^-$)   & 15.7(10) &10.2 & 16.2(5) \\
  B(E2;$\frac{7}{2}^-$)   & 3.4      & 5.1 &  9.92(14) \\
  B(E2;$\frac{5}{2}^-_1$) & -        & 1.5 &          \\ 
  B(E2;$\frac{5}{2}^-_2$) & -    & $\textless$ 0.1 &    \\[5pt]
  B(M1;$\frac{1}{2}^- $)   & 4.92(25) & 3.89(2) & 4.92(7)\\
  B(M1;$\frac{5}{2}^-_1$) & -        & 0.002(1)& \\
  B(M1;$\frac{5}{2}^-_2$) & -        & 0.02(1) & \\
\end{tabular}
\end{ruledtabular}
\caption{
  Selected $^7$Li observables calculated up through $N_{\max}=14$,
  with the same units as in Table~\ref{Tab:6Li}. 
  The energies are obtained from extrapolations to the infinite basis
  space, and the magnetic dipole observables are nearly converged,
  with error estimates as discussed in the text;
  the RMS point-proton radius and electric quadrupole observables 
  are evaluated at $\hbar\Omega=12.5$~MeV.
  Experimental values are from Refs.~\cite{Nortershauser:2011zz,F19881,Tilley20023}.
  AV18/IL2 results are from Refs.~\cite{Marcucci:2008mg,Pervin:2007,Pieper:2004qw,Pieper:2001ap}
  and include meson-exchange corrections for the dipole observables;
  CD-Bonn and INOY results are from Ref.~\cite{CDBINOY}, and were
  calculated at $N_{max}$=12 and $\hbar\Omega$=11 and 16 MeV
  respectively for CD-Bonn and INOY, with the INOY gs energy
  extrapolated to the infinite basis space.
  \label{Tab:7Li}}
\end{table}

Our results with JISP16 for selected spectral and other observables of
$^7$Li are summarized in Table~\ref{Tab:7Li} and compared with
experiment when available.  The error estimates for the excitation
energies are calculated as discussed above.  From this table we see
that the gs energy is underbound by about 0.67~MeV.  The gs energy and
excitation energies compare favorably to other methods and
interactions~\cite{Pieper:2004qw,Nogga:2005hp}.
For the RMS radius and the quadrupole observables (which are not
converged) we list our results from the largest basis space,
$N_{\max}=14$, at $\hbar\Omega=12.5$~MeV; they are in reasonable
agreement with the available experimental data, given the basis space
limitations and long-range nature of these operators.

\begin{figure}[t]
\includegraphics[width=0.95\columnwidth]{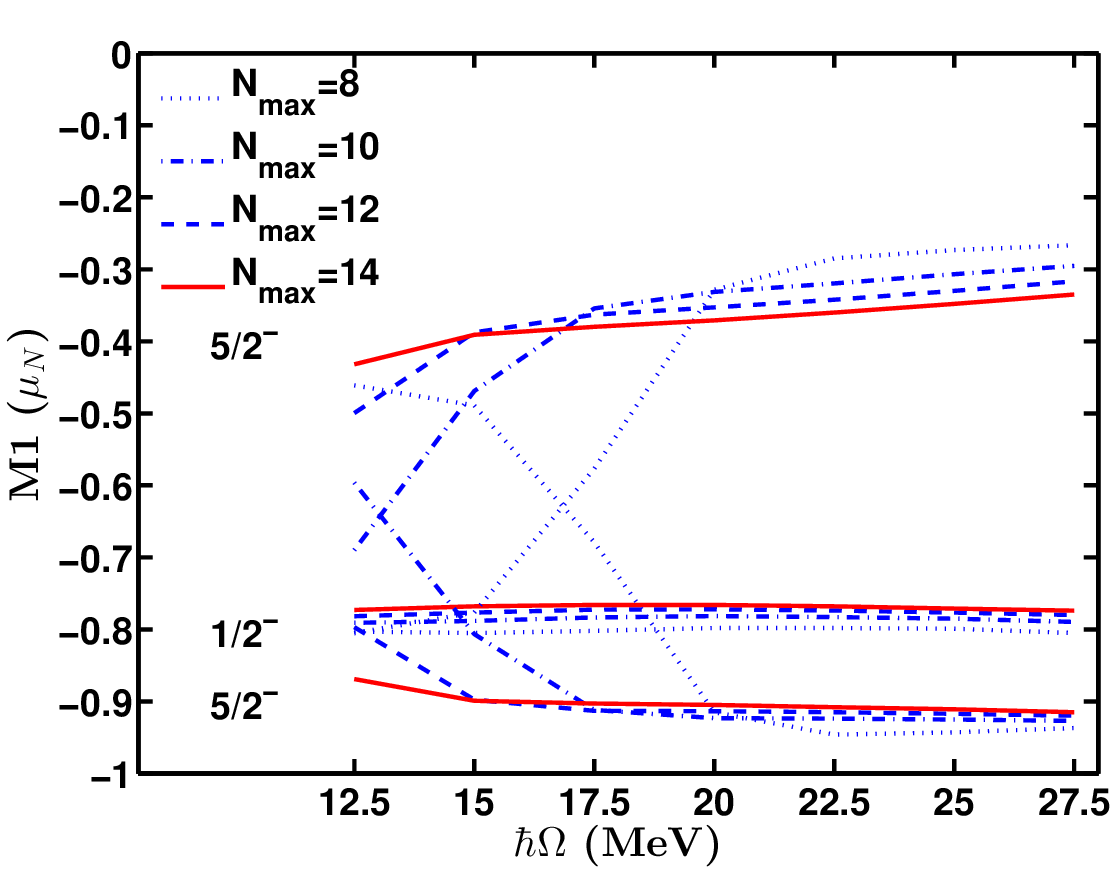}
\includegraphics[width=0.95\columnwidth]{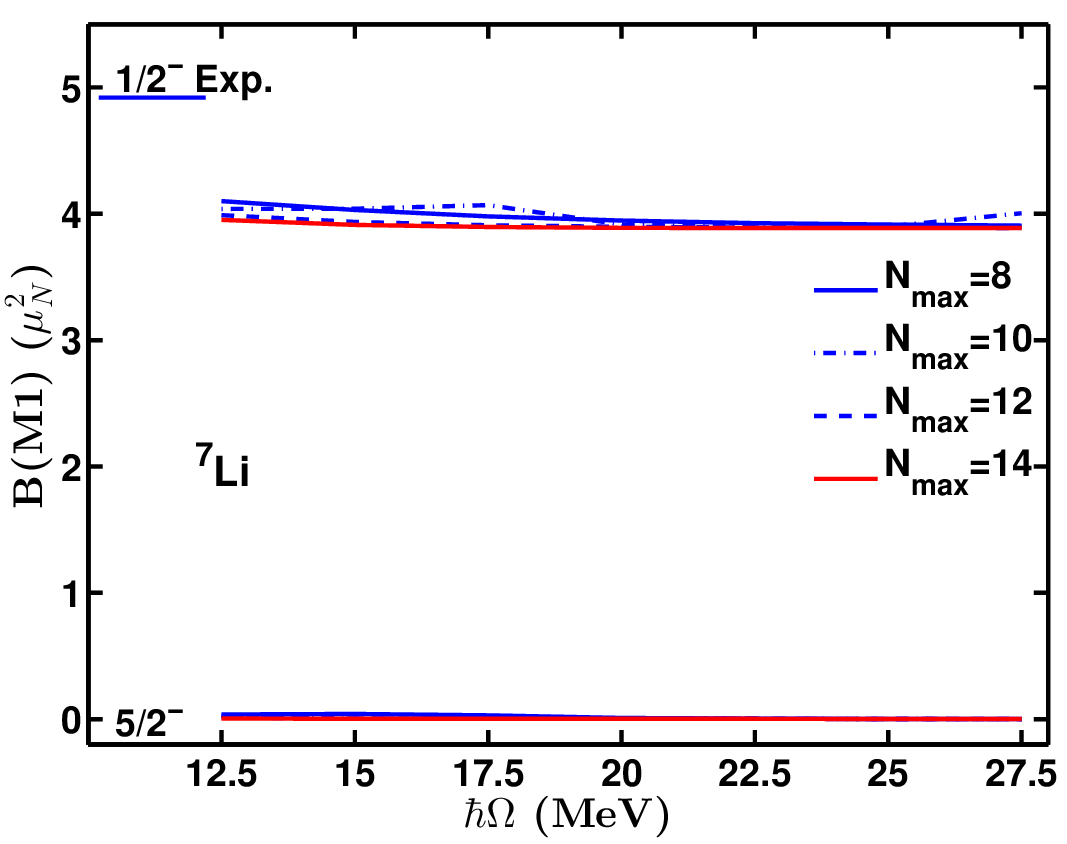}
\caption{(Color online)
  Dipole moment (top) and B(M1) transition to the gs (bottom) as a
  function of the basis $\hbar\Omega$ for selected excited states 
  of $^7$Li at various $N_{\max}$ values as indicated in the legend.  
  The experimental value for the B(M1) transition from the excited
  $\frac{1}{2}^-$ to the ground state is indicated on the left side 
  of the lower panel and is about 20\% larger than the theoretical
  result.
  \label{Fig:magnetic7Li}}
\end{figure}

Fig.~\ref{Fig:magnetic7Li} displays the magnetic dipole moments and
reduced magnetic dipole transition probabilities to the gs for selected states of
$^7$Li.  These observables converge quickly, and for calculations at
$N_{\max}=14$, are almost independent the HO energy.  In fact, most of
the magnetic observables are already reasonably well converged (to
within 10\%) at $N_{\max}=8$, with the noticable exception of the two
$\frac{5}{2}^-$ excited states.  This is partially due to a strong
state mixing between these two states.  We require larger basis spaces
to fully differentiate these states, because they are close together
in energy, and their quantum numbers are identical.

Our estimate for the infinite basis space results for magnetic dipole
observables is based on the residual dependence on $N_{\max}$ and
$\hbar\Omega$ over a 10 MeV window in $\hbar\Omega$.  This window does
include the optimal $\hbar\Omega$ for the extrapolations and the
variational upperbound, but is not necessarily centered around these
values.  Our numerical error estimate is the RMS sum of the variation
with $\hbar\Omega$ over this window and the difference between the
results in the two largests $N_{\max}$ calculations (rounded up),
i.e. treating the variation with each of the two basis space
parameters as independent sources of numerical uncertainties.

With JISP16, the magnetic moment of the gs is about 10\% too low
compared to experiment.  This could easily arise from our neglect of
meson-exchange currents in our current calculations.  GFMC
calculations with AV18 plus Illinois-2 three-body
forces~\cite{Marcucci:2008mg} found that the magnetic moment of the
$^7$Li gs receives a 10\% correction from meson-exchange currents,
changing the magnetic moment from $2.9~\mu_N$ to $3.2~\mu_N$.  It is
quite remarkable that our result for the magnetic moment, with the
naive pointlike M1 operator, is in fact quite close to the 
results
obtained with CD-Bonn, with INOY, and with AV18 plus Illinois-2, all 
about 10\% below the experimental datum.  
Apparently, this observable is not very sensitive 
to the details of the interaction.
(Note that the exchange
current correction to the $^6$Li gs magnetic moment was only 2\% in
Ref.~\cite{Marcucci:2008mg}.)

The B(M1) from the $\frac{1}{2}^-$ to the gs is about 20\% too low
compared to experiment.  Again, this is in qualitative agreement with
the findings of Ref.~\cite{Marcucci:2008mg}: with AV18 plus Illinois-2
there is about 10\% increase in the M1 transition matrix element due
meson-exchange currents, which results in a 20\% increase in the
corresponding B(M1).

\begin{figure}[t]
\includegraphics[width=0.95\columnwidth]{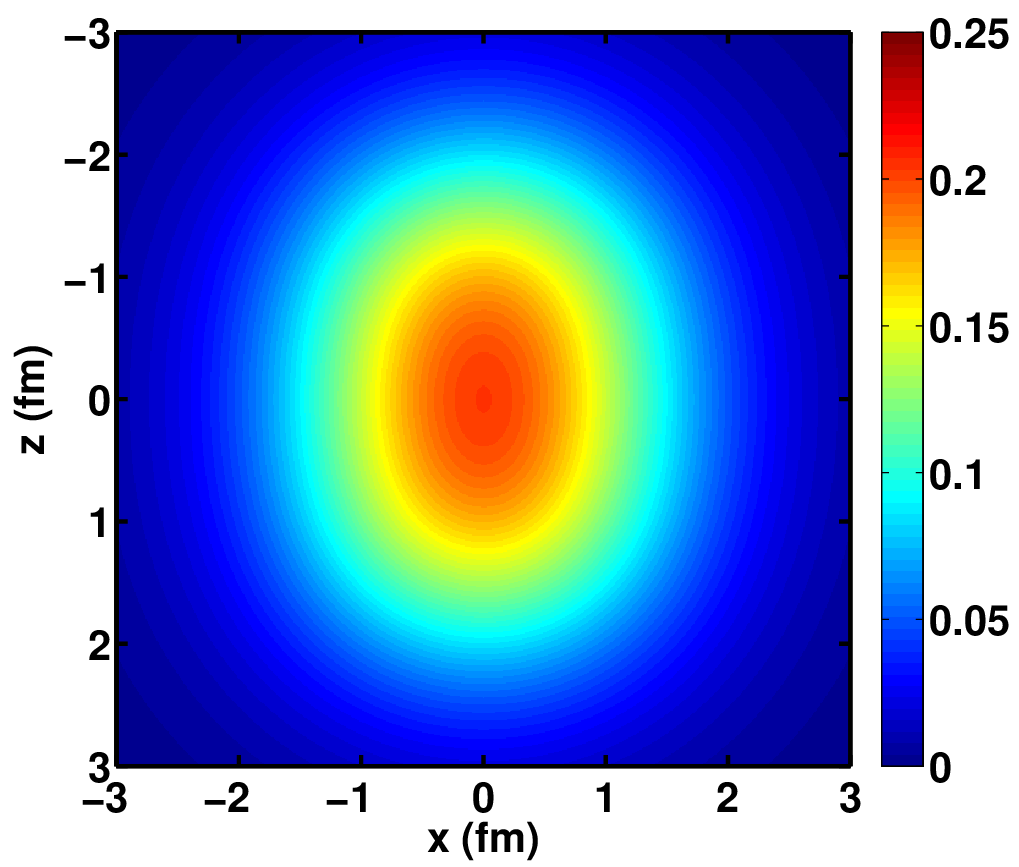}
\includegraphics[width=0.95\columnwidth]{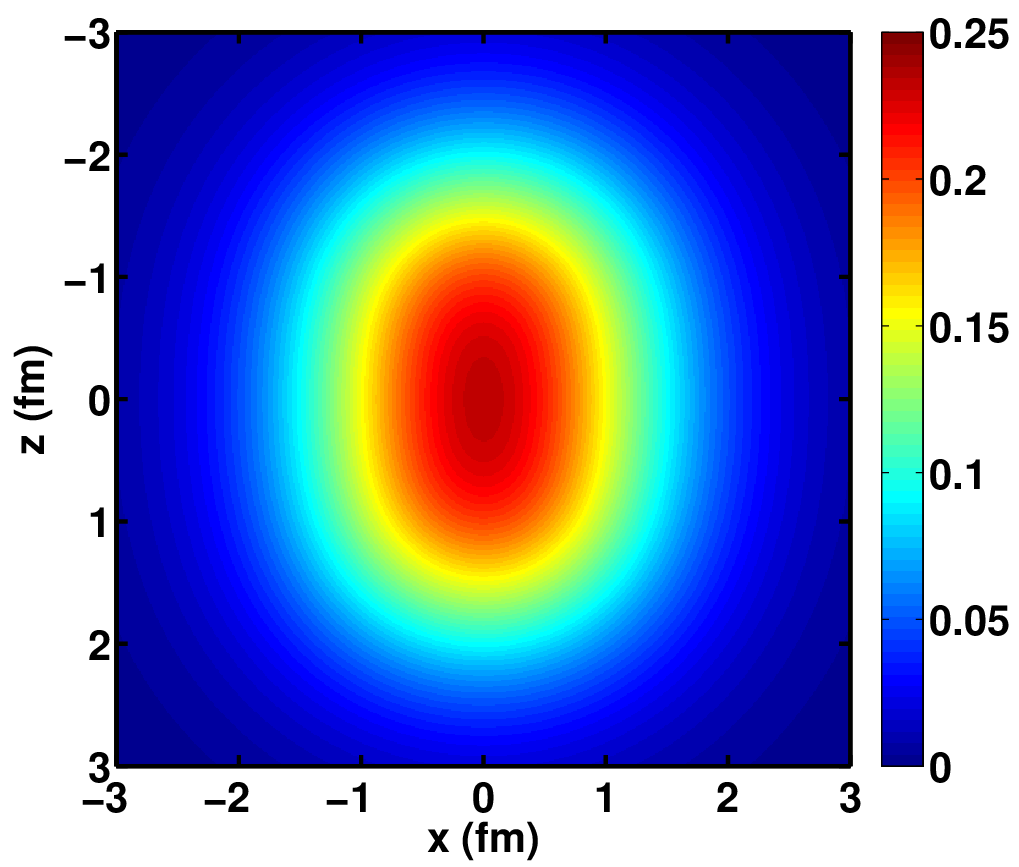}
\caption{(Color online)
  The $y=0$ slice of the gs matter density of $^7$Li
  before (top) and after (bottom) deconvolution of the spurious cm motion.
  These densities were calculated at $N_{\max}=14$ and $\hbar\Omega=12.5$~MeV.
  \label{Fig:spuriousdens3D7Li}}
\end{figure}

The effect of the cm motion on the density is shown in
Fig.~\ref{Fig:spuriousdens3D7Li} for the gs of $^7$Li.  The top panel
shows the space-fixed (sf) density including the cm motion,
$\rho^\Omega_{\rm sf}(\vec{r})$, whereas the bottom panel shows the
translationally-invariant density, $\rho_{\rm ti}(\vec{r})$.  The
smearing of the density due to the cm motion leads to a diminished
central density; the sf density has a central value of 0.204
nucleons/fm$^3$ while the ti density has a central value of 0.233
nucleons/fm$^3$.

The cm motion smearing spreads out the sf density leading to a slower
falloff and a larger radius than the ti density.  Furthermore, the ti
density has a more pronounced oblate shape than the sf density, as
would be expected from smearing with a spherically-symmetric function
that averages out the non-spherical details.  In order to characterize
the degree of deformation, we compare the ratio of the long axis to
the short axis of the elliptical density slices.  The ratio of the
long to short axes at half central density is 1.78 for the ti density
and 1.60 for the sf density.  Note that the extent of the smearing
effect from cm motion depends on the HO energy of the basis.  The sf
density depends on $\hbar\Omega$, even in the limit $N_{\max} \to
\infty$, whereas the ti density becomes independent of the basis in
this limit.

\begin{figure}[t]
\includegraphics[width=0.95\columnwidth]{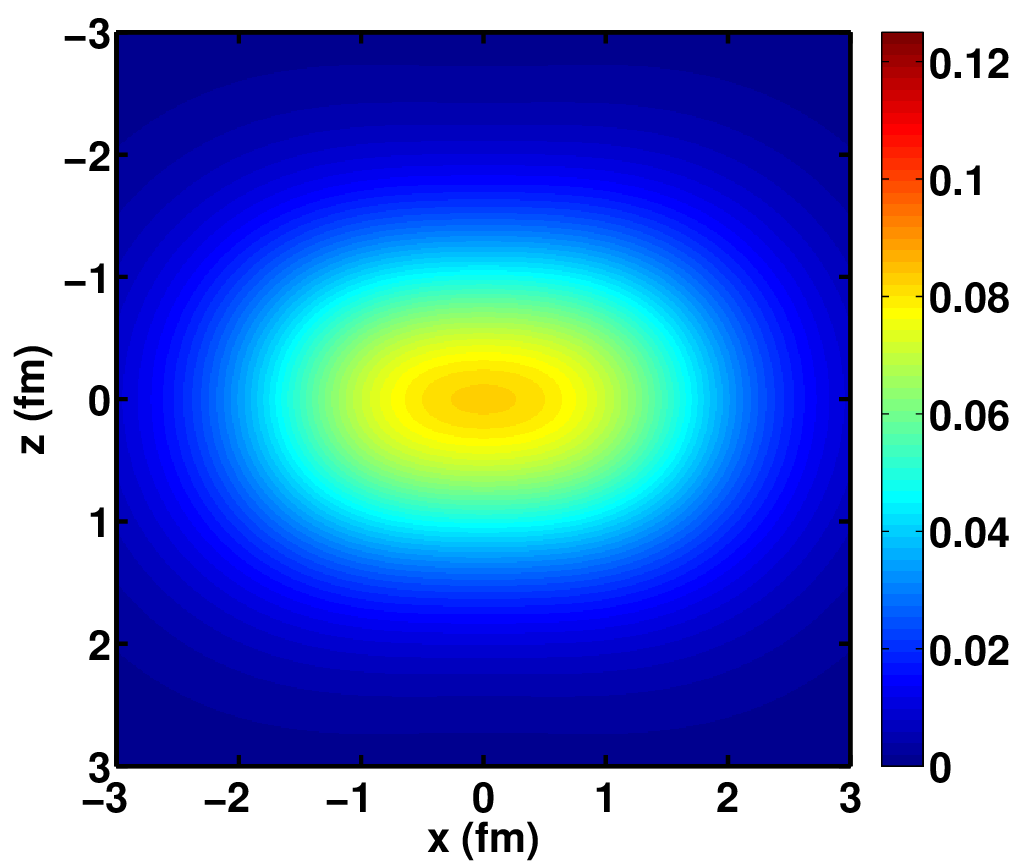}
\includegraphics[width=0.95\columnwidth]{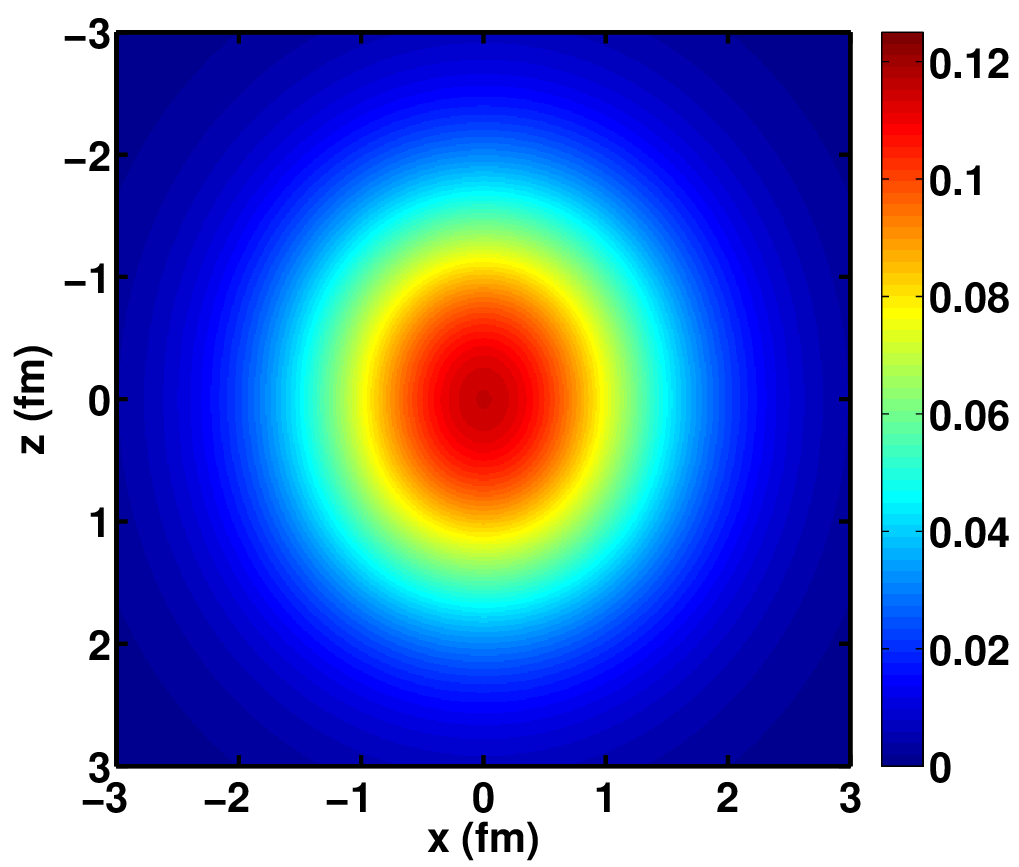}
\caption{(Color online)
  The $y=0$ slices of the translationally-invariant proton densities
  for the first excited $\frac{5}{2}^-$ state (top) and for the second
  excited $\frac{5}{2}^-$ state (bottom) of $^7$Li. These densities
  were calculated at $N_{\max}=14$ and $\hbar\Omega=12.5$~MeV.
  \label{Fig:dens3D7Li5}}
\end{figure}

Fig.~\ref{Fig:dens3D7Li5} contrasts proton densities of the fourth and
fifth excited states of $^7$Li.  Although their quantum numbers are
equal, $(J^\pi, T) = (\frac{5}{2}^-,\frac{1}{2})$, they have other
features that make them quite distinct.  Experimentally, the first
excited $\frac{5}{2}^-$ is broad, whereas the second excited
$\frac{5}{2}^-$ is narrow: their widths are $0.88$~MeV and $0.09$~MeV
respectively.  Indeed, our calculated
excitation energy is better converged for the higher of
these two states.  Furthermore, our calculations show significant
differences in their structure: the first excited $\frac{5}{2}^-$ has
a large negative quadrupole moment, whereas the second has a moderate
positive quadrupole moment (see Table~\ref{Tab:7Li}).  Indeed, the
density in the top panel of Fig.~\ref{Fig:dens3D7Li5} is strongly
oblate, whereas the bottom panel shows a moderately prolate shape (we
note again that the densities are symmetric around the azimuthal axis,
which is the vertical axis is these plots).  Another noteworthy
difference is observed in the magnitude of the central proton density:
the more diffuse $(\frac{5}{2}^-,\frac{1}{2})_1$ state has a central
proton density of only 0.08 protons/fm$^3$ while the
$(\frac{5}{2}^-,\frac{1}{2})_2$ state has a central proton density of
0.12 protons/fm$^3$, 50\% higher.

\subsection{$^8$Li}

\begin{table}[t]
\renewcommand{\arraystretch}{1.2} 
\begin{ruledtabular}
\begin{tabular}{lcrrrr}
  $^8$Li          & Expt. & JISP16   & AV18/IL2 & CD-B & INOY\\ \hline
  $E_{\rm b}(2^+)$  & 41.277 & 40.3(2) & 41.9(2)  & 35.82   & 41.3(5) \\
  $\langle r^2_{pp}\rangle^{1/2}$ & 2.21(6) & 2.1 & 2.09(1) & 2.17 & 2.01 \\[5pt]
  $E_{\rm x}(1^+)$     & 0.981      & 1.5(2)    &   1.4(3) \\ 
  $E_{\rm x}(3^+)$     & 2.255(3)   & 2.8(1)    &   2.5(3) \\ 
  $E_{\rm x}(4^+)$     & 6.53(2)    &  7.0(3)   &   7.2(3) \\[5pt] 
  $Q(2^+)$            & 3.27(6)   &  2.6   & 3.2(1) & 2.78 & 2.55 \\
  $Q(1^+)$            & -         &  1.2   & \\ 
  $Q(3^+)$            & -         & -2.0   & \\
  $Q(4^+)$            & -         & -3.4   & \\[5pt]
  $\mu$($2^+$)        & 1.654    &  1.3(1)  & 1.65(1) & 1.24 & 1.42 \\
  $\mu$($1^+$)        & -        & -2.2(2)  & \\
  $\mu$($3^+$)        & -        &  2.0(1)  & \\
  $\mu$($4^+$)        & -        &  1.84(1) & \\[5pt]
  B(E2;$1^+$) & -     & 1.9 & \\
  B(E2;$3^+$) & -     & 4.6 & \\
  B(E2;$4^+$) & -     & 1.9 & \\[5pt]
  B(M1;$1^+$) & 5.0(16)     & 3.7(2)   & \\
  B(M1;$3^+$) & 0.52(23)    & 0.25(5)  & \\
\end{tabular}
\end{ruledtabular}
\caption{
  Selected $^8$Li observables calculated up through $N_{\max}=12$,
  with the same units as in Table~\ref{Tab:6Li}. 
  The energies are obtained from extrapolations to the infinite basis
  space, and the magnetic dipole observables are nearly converged,
  with error estimates as discussed in the text;
  the RMS point-proton radius and electric quadrupole observables 
  are evaluated at $\hbar\Omega=12.5$~MeV.
  Experimental values are from Refs.~\cite{Nortershauser:2011zz,Tilley2004155,Audi20033}.
  AV18/IL2 results are from Refs.~\cite{Pieper:2001ap,Pieper:2004qw}
  and does not include meson-exchange corrections for the magnetic moment;
  CD-Bonn and INOY resuts are from Ref.~\cite{CDBINOY}, and were
  calculated at $N_{max}$=12 and $\hbar\Omega$=12 and 16 MeV
  respectively for CD-Bonn and INOY, with the INOY gs energy
  extrapolated to the infinite basis space.
  \label{Tab:8Li}}
\end{table}

\begin{figure}[t]
\includegraphics[width=0.95\columnwidth]{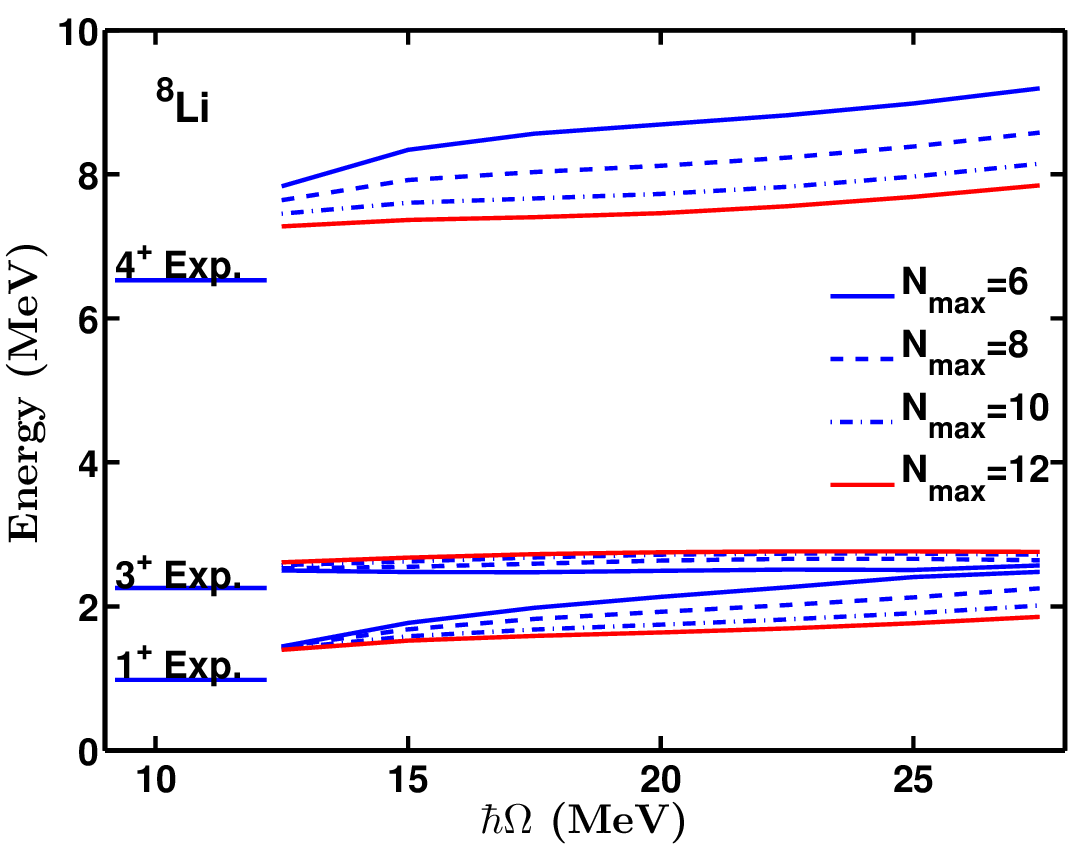}
\caption{(Color online)
  Excitation energies for select excited states of $^8$Li are
  shown as a function of $\hbar\Omega$ at various $N_{\max}$ values as
  indicated in the legend.
  \label{Fig:energy8Li}}
\end{figure}

Table~\ref{Tab:8Li} presents a capsule view of selected spectral and
other observables for $^8$Li calculated with JISP16 and compared with
experiment when available.  The error estimates for the excitation
energies and magnetic dipole observables are calculated as discussed
above.  For the RMS point-proton radius and the charge quadrupole
observables (which are not converged) we list our results from the
largest basis space, $N_{\max}=12$, at $\hbar\Omega=12.5$~MeV.  From
this table we see that the gs energy is underbound by about 1.0~MeV.

In addition to the gs, we examined several narrow low-lying states:
the lowest two excited states, which are narrow states, with a width
of 33 keV or less, as well as a narrow low-lying $4^+$ state at
$6.53$~MeV with a width of 35 keV.  We do not consider isospin $2$
states -- the lowest isospin $2$ has more than $10$ MeV excitation
energy.  The excitation energies obtained with JISP16 compare
reasonably well with the experimental excitation energies, though the
level splittings are a bit too large (see Fig.~\ref{Fig:energy8Li}).
The convergence of the spectrum is similar as for the other
Li-isotopes, though the convergence of the $4^+$ state is somewhat
slower than expected based on its small width.  In our calculations
there are several additional states below this $4^+$ state, which is
typically the 8th state in our calculated spectrum, depending on the
basis space; experimentally, it is the 7th observed state.

The convergence properties of the calculated magnetic dipole
observables are similar to those for $^6$Li and $^7$Li discussed
above.  Because we can only go up to $N_{\max}=12$, the numerical
error estimates are slightly larger than for $^6$Li and $^7$Li.  The
gs magnetic moment is approximately 20\% lower than experiment.  This
seems reasonable, in light of our discussions of magnetic moments
above, that this discrepancy is at least partially due to the fact
that we do not incorporate meson-exchange currents.  

Note that CD-Bonn
gives a similar magnetic moment as JISP16, but that INOY provides a
moment that is slightly closer to experiment, and AV18/IL2 gives a
magnetic moment in excellent agreement with data; however, the contributions
of meson-exchange currents were not evaluated for this magnetic moment.  

The reduced B(M1)
transition probabilities from the $1^+$ and $3^+$ to the gs are 20\% and 50\%
lower than experiment, but the experimental error bars are large.  The
magnetic moment of the $4^+$ state is remarkably well converged,
despite its excitation energy not being very well converged.

\begin{figure*}[t]
\includegraphics[width=0.95\columnwidth]{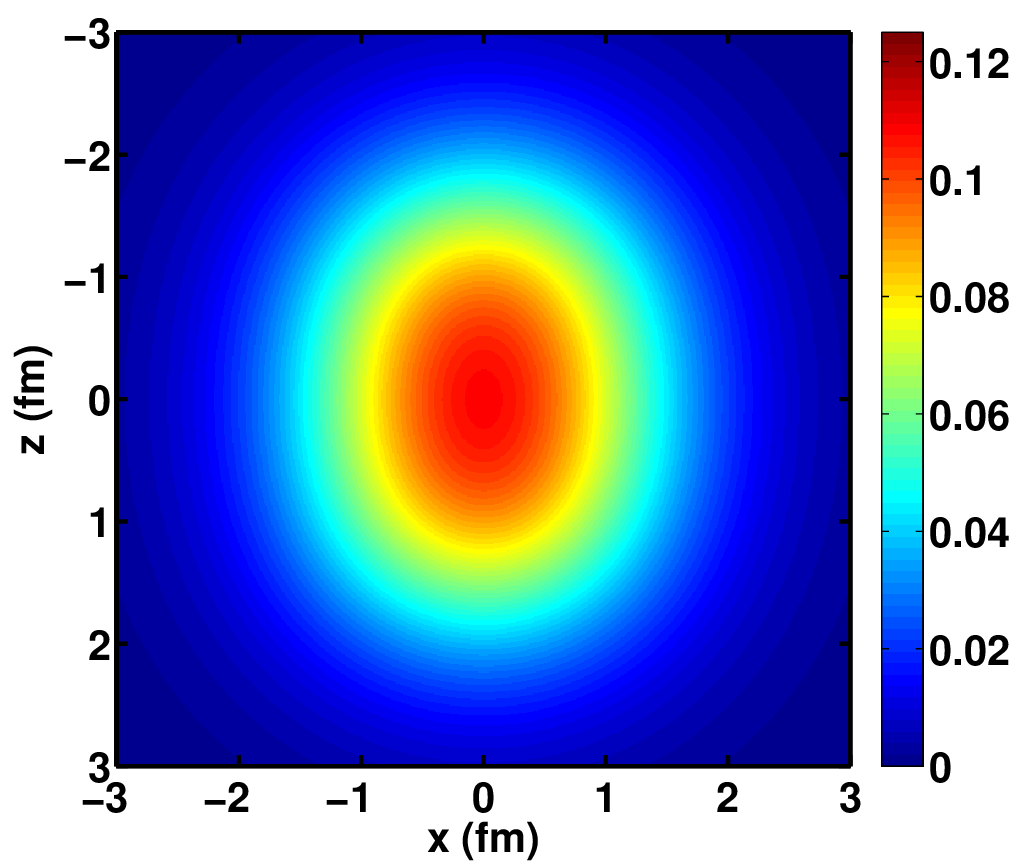}
\qquad\includegraphics[width=0.95\columnwidth]{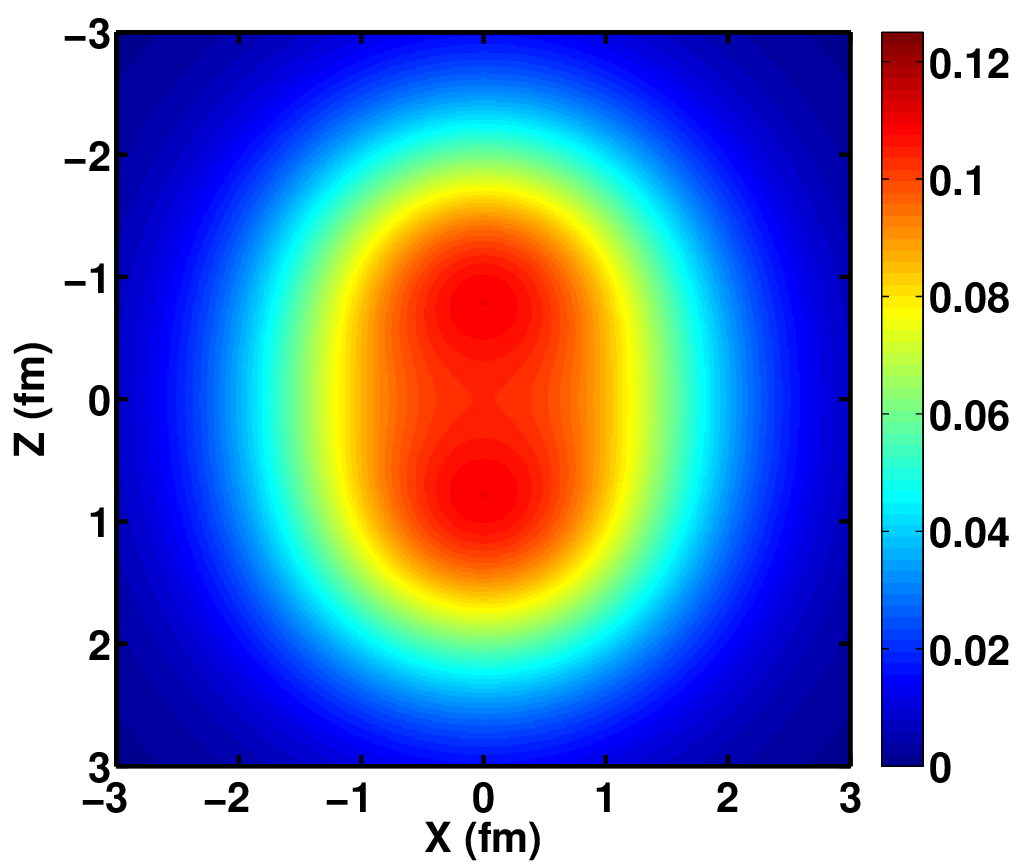}
\includegraphics[width=0.95\columnwidth]{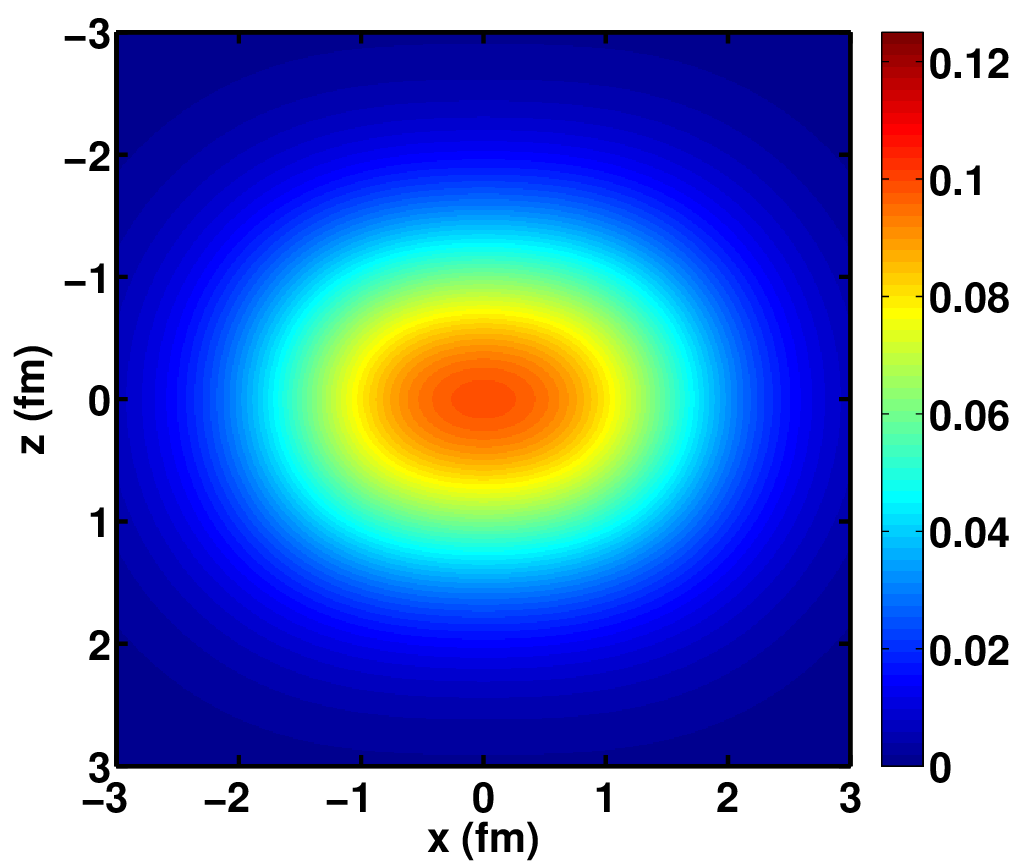}
\qquad\includegraphics[width=0.95\columnwidth]{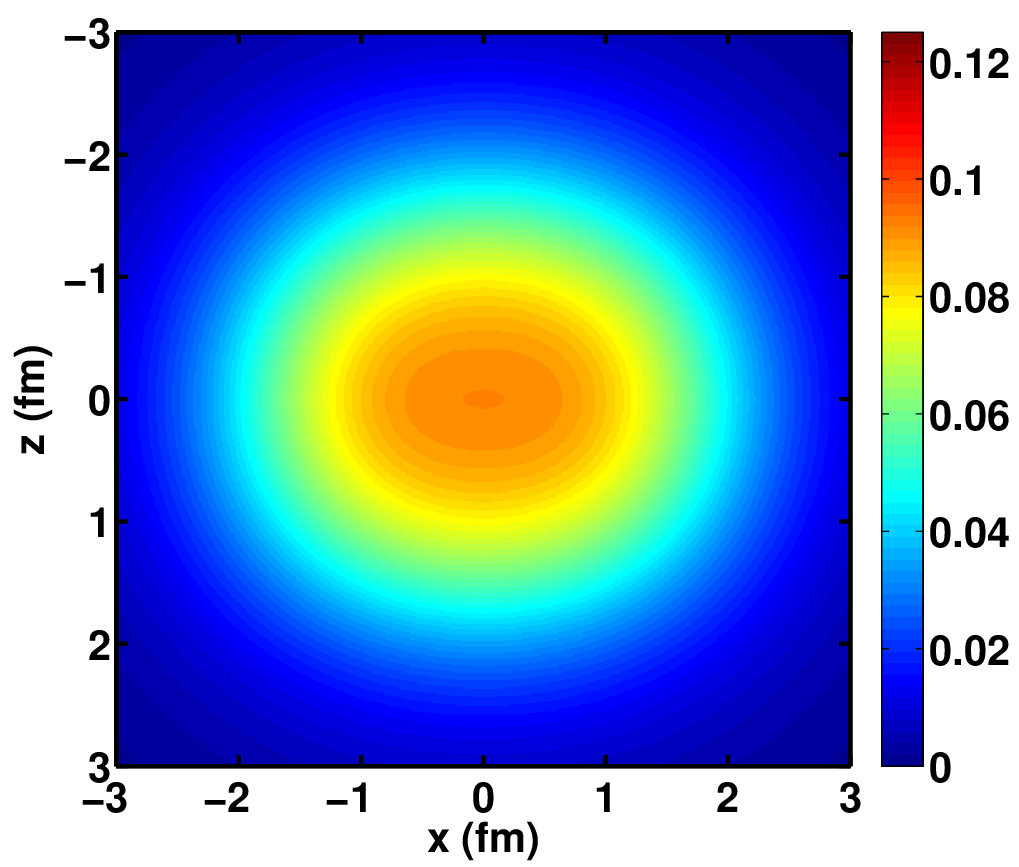}
\caption{(Color online)
  The $y=0$ slice of the translationally-invariant proton
  (left) and neutron (right) densities of the $2^+$ gs (top) and 
  the first excited $4^+$ state (bottom) of $^8$Li.  These densities were
  calculated at $N_{\max}=12$ and $\hbar\Omega=12.5$~MeV.
  \label{Fig:dens8Li}}
\end{figure*}
The quadrupole moments and reduced B(E2) transition probabilities are not well
converged due to basis space limitations, as discussed above.  In
spite of these limitations, the electric quadrupole moments allow us to
qualitatively understand the shape of the proton densities of these
states: prolate for the $2^+$ gs and the first excited $1^+$, but
oblate for the first excited $3^+$ and $4^+$ states.  Indeed, that is
what we see in the left-hand panels of Fig.~\ref{Fig:dens8Li}, where
we plot these densities for the gs and for the $4^+$ state.

\begin{figure*}[t]
\includegraphics[width=0.95\columnwidth]{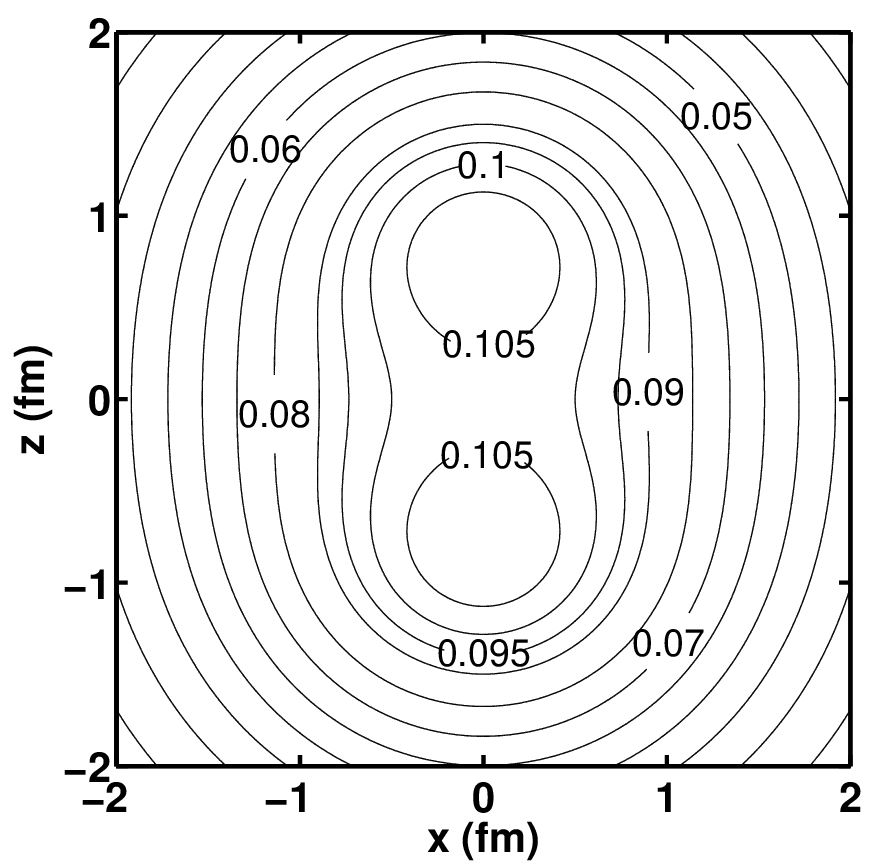}\qquad
\includegraphics[width=0.95\columnwidth]{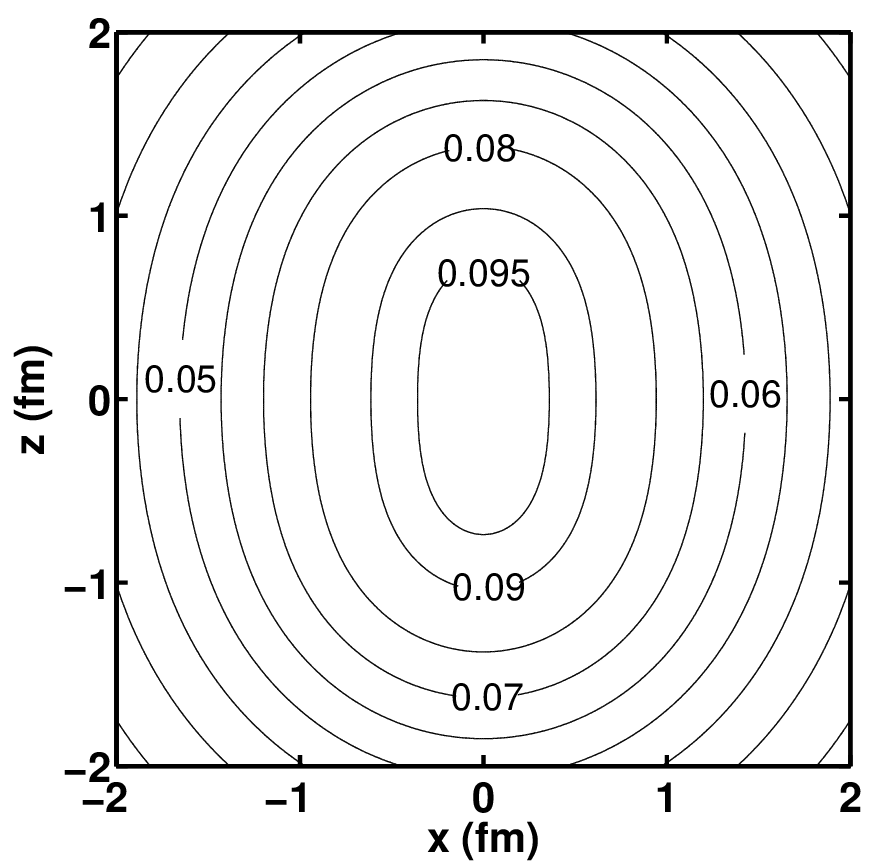}
\caption{(Color online) The $y=0$ slice of the
  translationally-invariant neutron density (left) of the $2^+$ gs of
  $^8$Li. The space-fixed neutron density for the same state is on the
  right. The contour labels give the density in neutrons/fm$^3$.
  These densities were calculated at $N_{\max}=12$ and
  $\hbar\Omega=12.5$~MeV.
\label{Fig:dens8LiZoom}}
\end{figure*}

Interestingly, the neutron density differs by more than a simple scale
change from the proton density for these two states, as can be seen
from the right-hand panels of Fig.~\ref{Fig:dens8Li}.  In the $2^+$
state, the deformation of the neutrons is significantly larger than
that of the protons, whereas in the $4^+$ state, the deformation of
the neutrons is much smaller than that of the protons.  

A case of special interest can be seen in the top right panel of
Fig.~\ref{Fig:dens8Li}, or in more detail in
Fig.~\ref{Fig:dens8LiZoom}.  In the left panel of
Fig.~\ref{Fig:dens8LiZoom} we clearly see non-trivial neutron
clustering that is obfuscated in the sf frame (right), highlighting
the importance of the deconvolution procedure and the significance of
the ti density.
Furthermore, the ti density (left panel) has a significantly higher
density in the central region than the sf density (right panel).  Both
the ti and the sf densities are normalized to give the same integrated
density of five neutrons, so that means that the sf density is smeared
out over a larger region, and falls off to zero slower than the ti
density.  This is also evident in Fig.~\ref{Fig:spuriousdens3D7Li},
where we contrasted the sf and ti matter density of the gs of $^7$Li:
the ti central density is significantly higher than the sf central density.

\begin{figure*}[t]
\includegraphics[width=0.95\columnwidth]{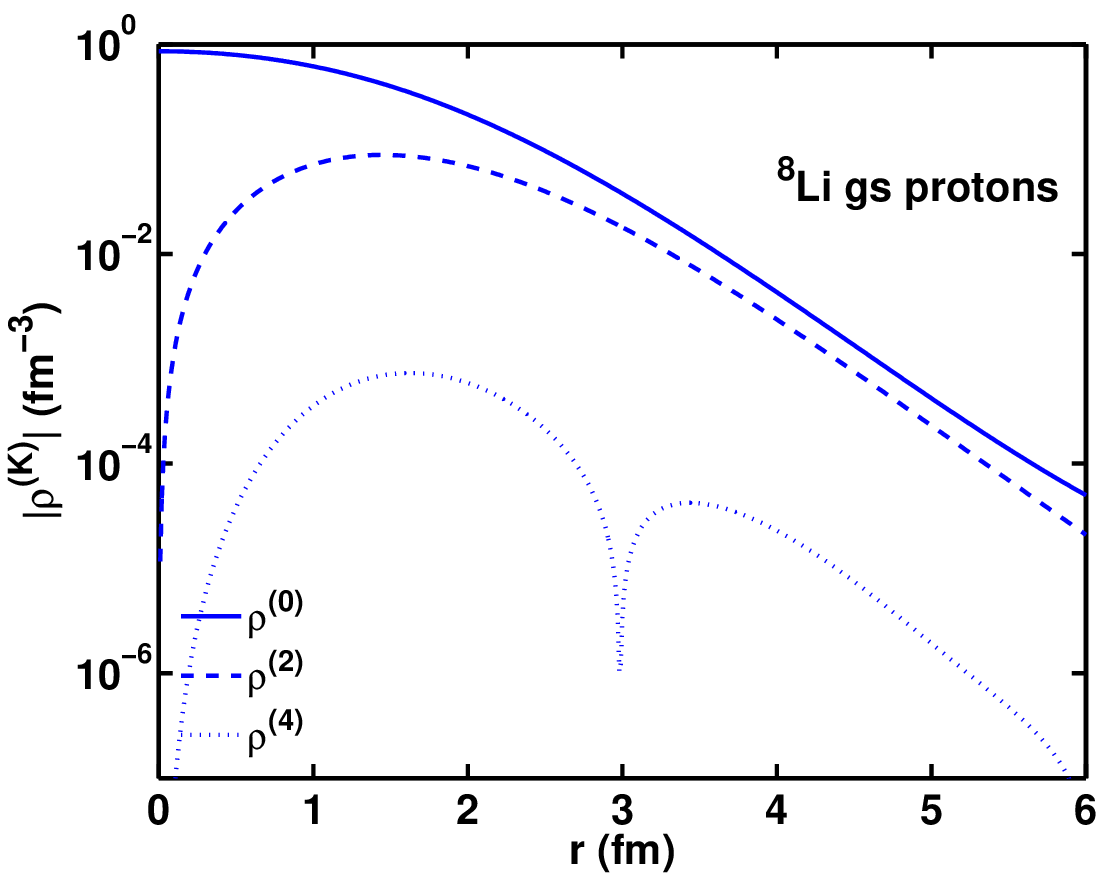}
\qquad\includegraphics[width=0.95\columnwidth]{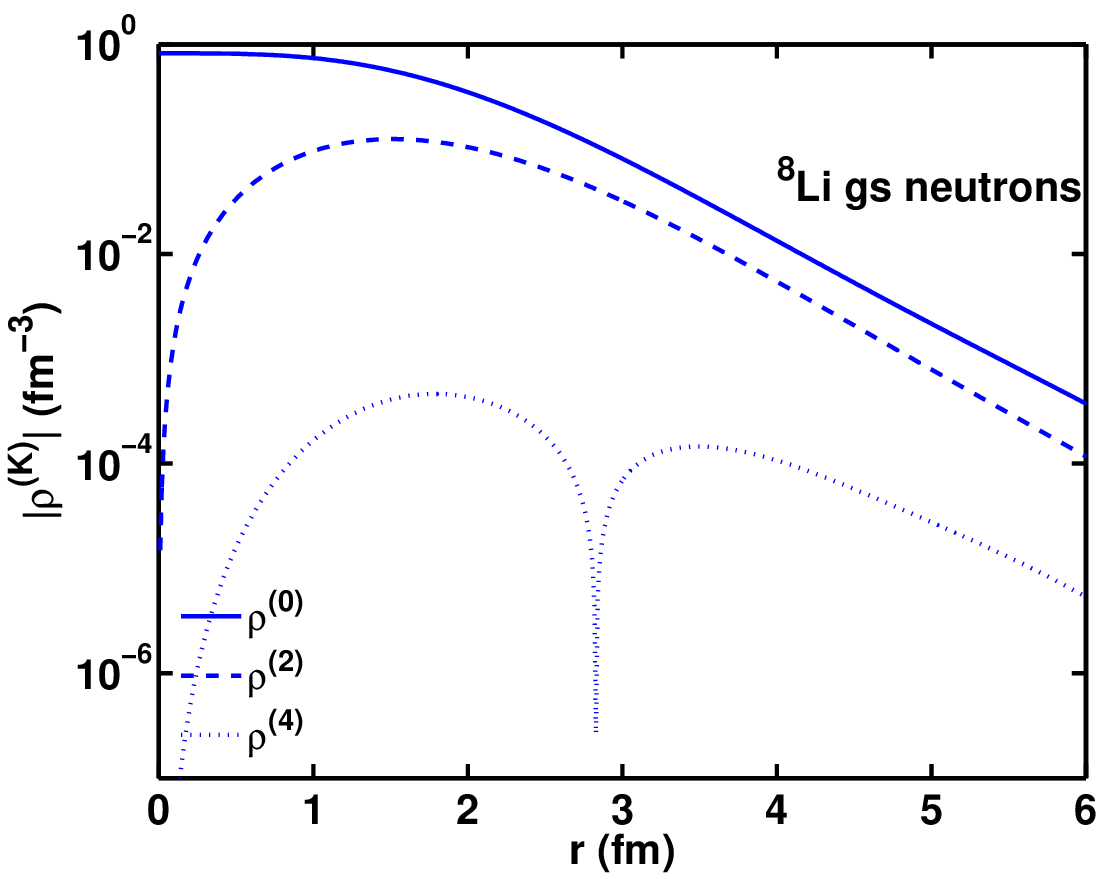}
\includegraphics[width=0.95\columnwidth]{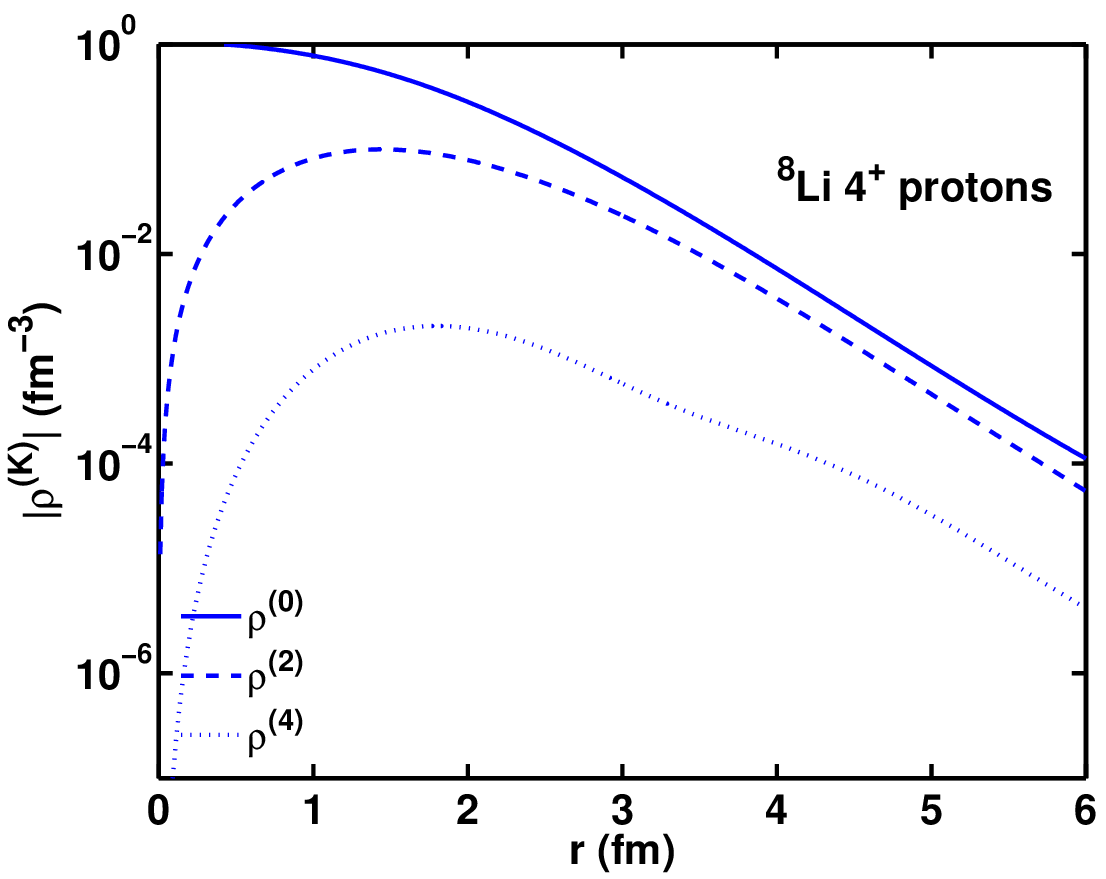}
\qquad\includegraphics[width=0.95\columnwidth]{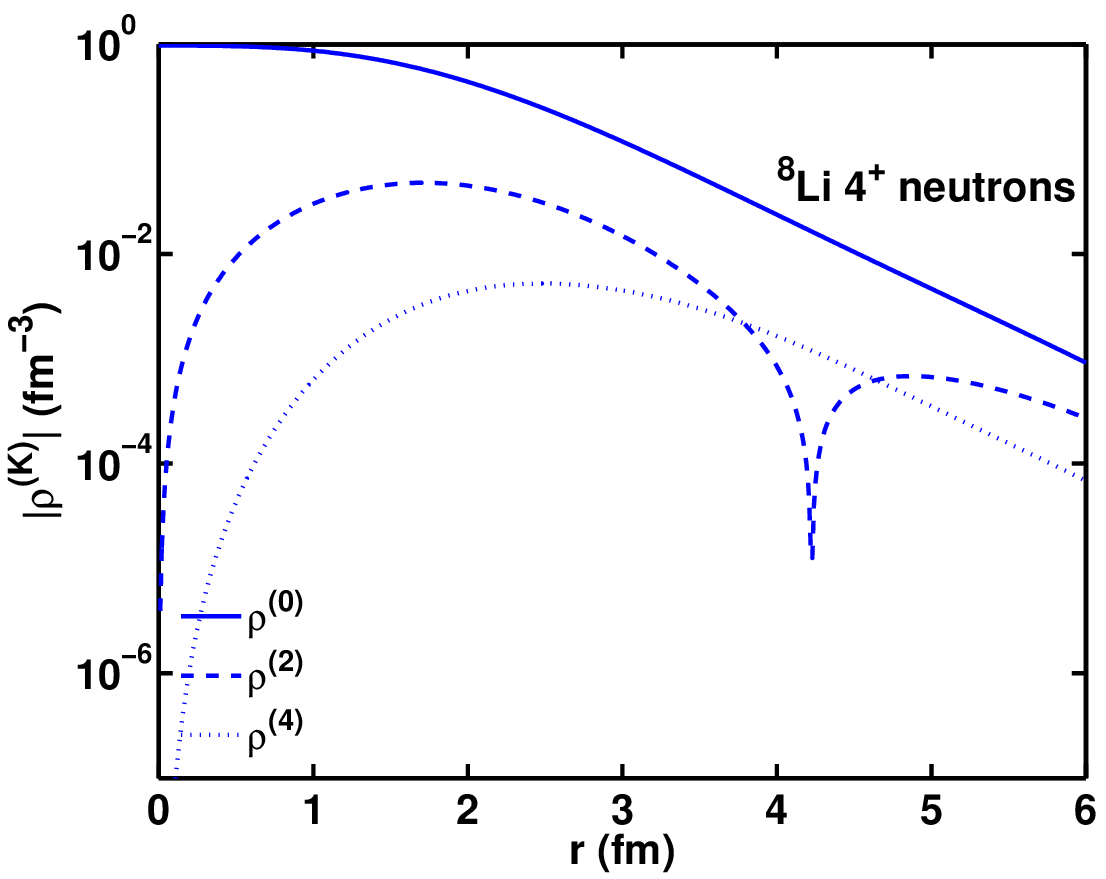}
\caption{(Color online) 
  The multipole components $\rho^{(K)}_{ti}(r)$ of the proton (left) and
  neutron (right) densities of the $2^+$ gs (top) and the first
  excited $4^+$ state (bottom) of $^8$Li.  These densities were calculated at
  $N_{\max}=12$ and $\hbar\Omega=12.5$~MeV.
Monopole and quadrupole distributions for the gs are all positive. The $K=4$ distributions for the gs are negative in the interior and positive in the tail region.  For the $4^+$ state, the monopoles are positive while the quadrupole is negative for the protons and negative for the interior of the neutrons.  Both $K=4$ distributions are positive for the $4^+$ state.
\label{Fig:dens8LiMultiPole}}
\end{figure*}

\begin{figure*}[t]
\includegraphics[width=0.95\columnwidth]{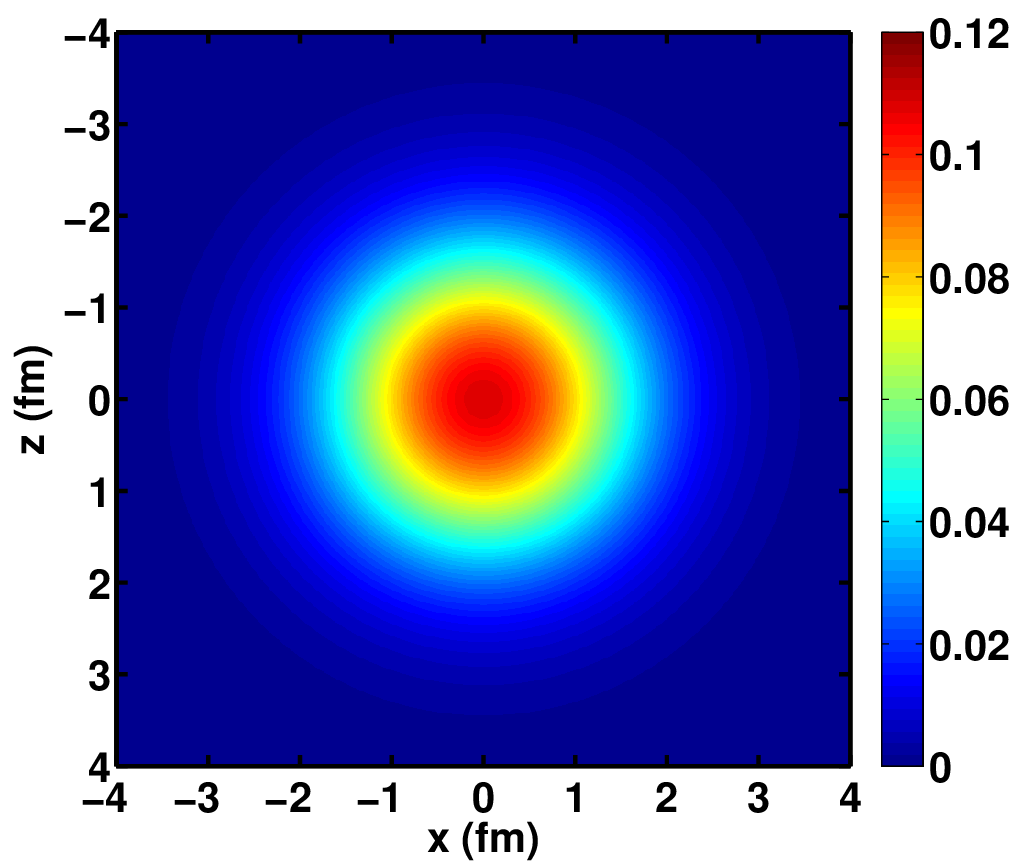}
\qquad\includegraphics[width=0.95\columnwidth]{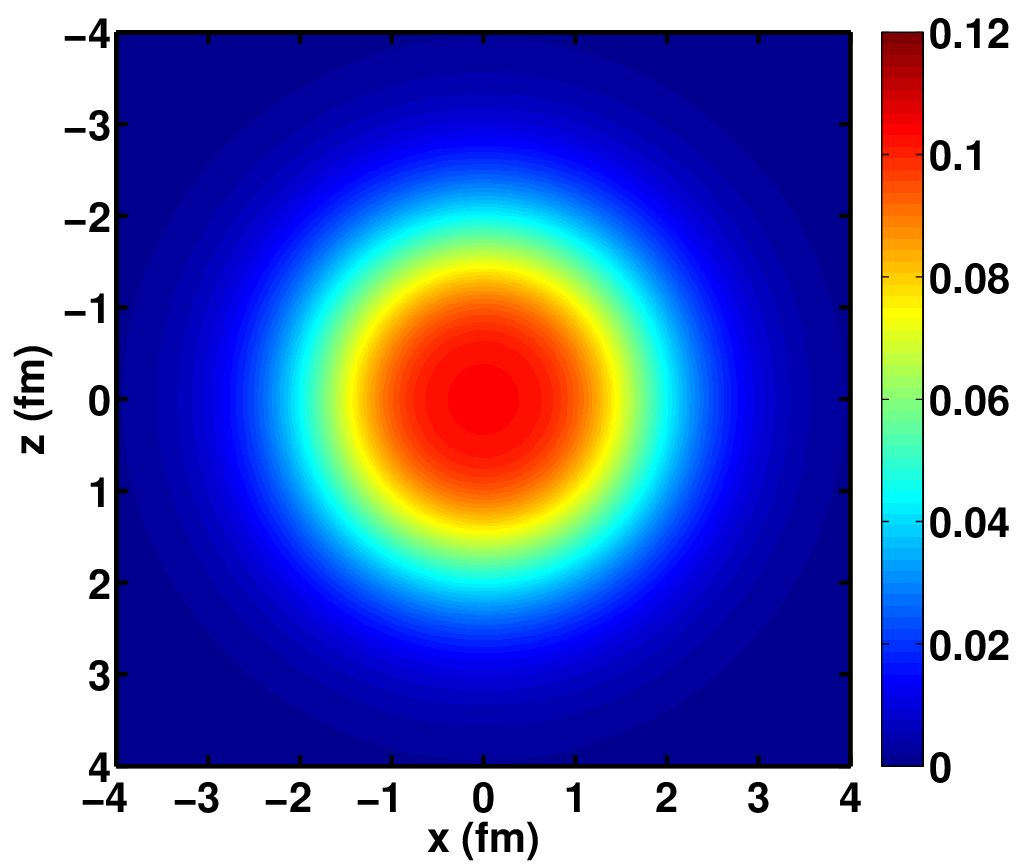}
\includegraphics[width=0.95\columnwidth]{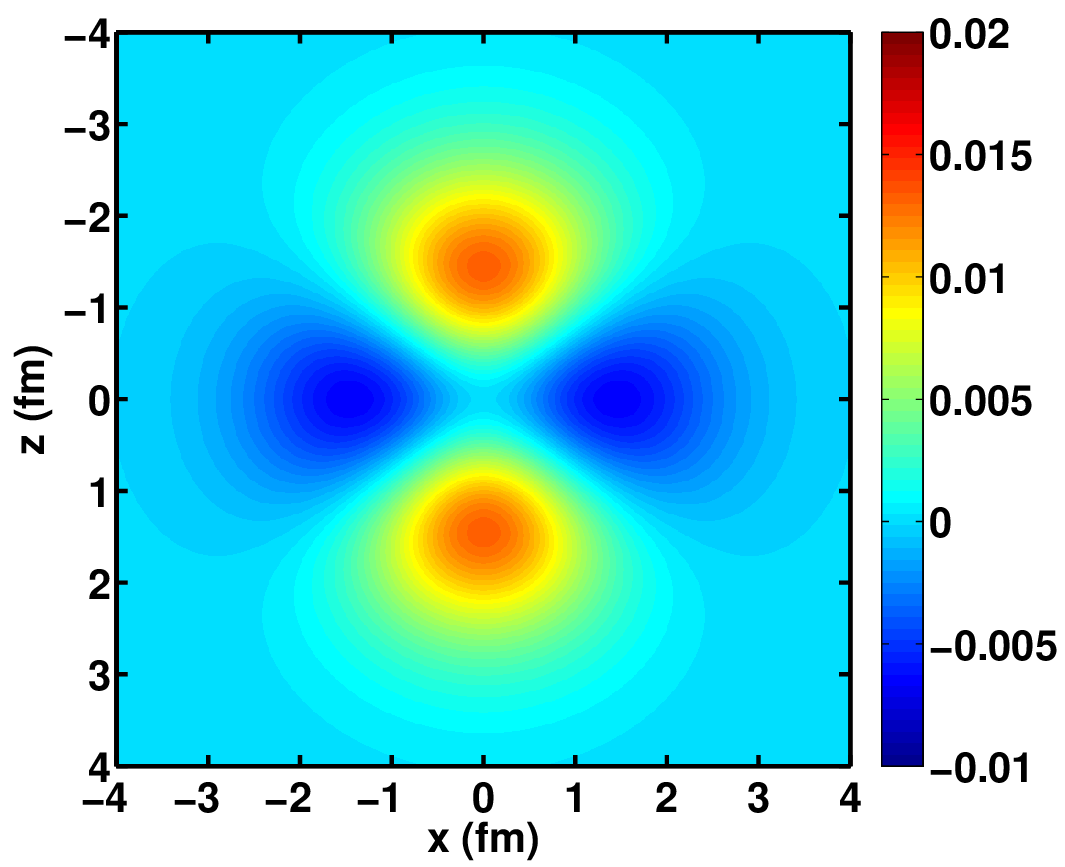}
\qquad\includegraphics[width=0.95\columnwidth]{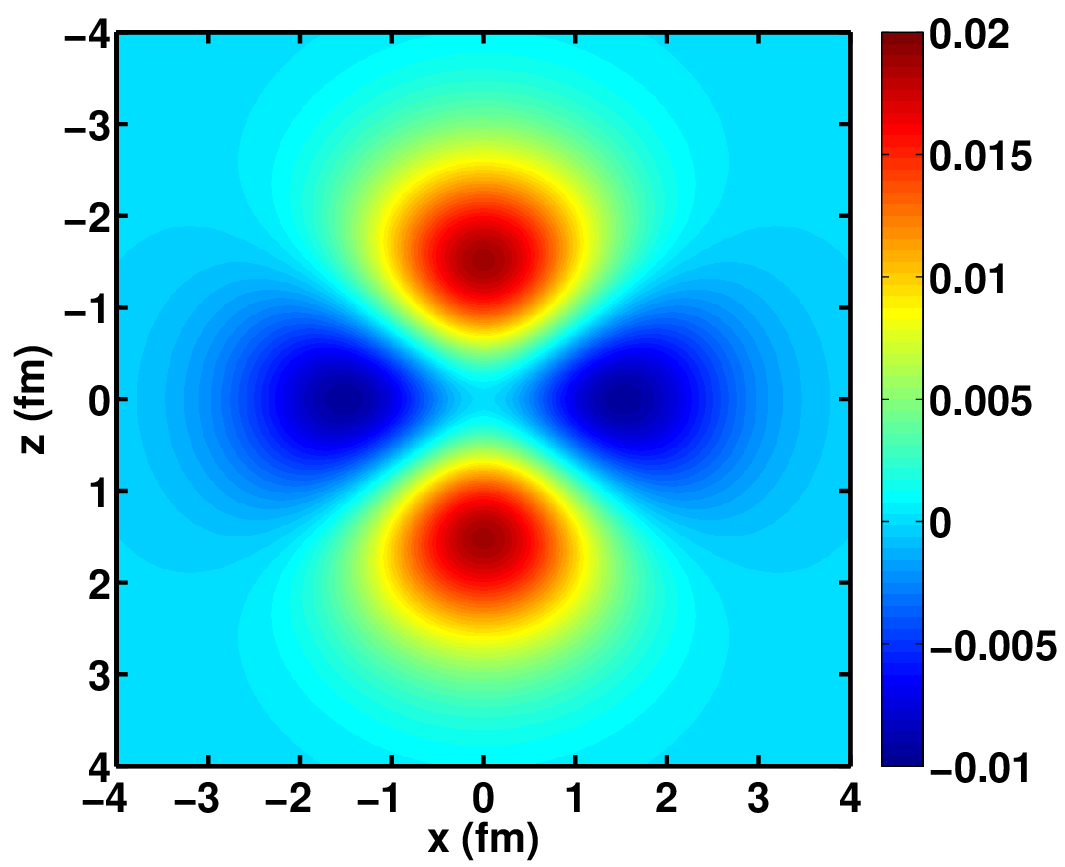}
\includegraphics[width=0.95\columnwidth]{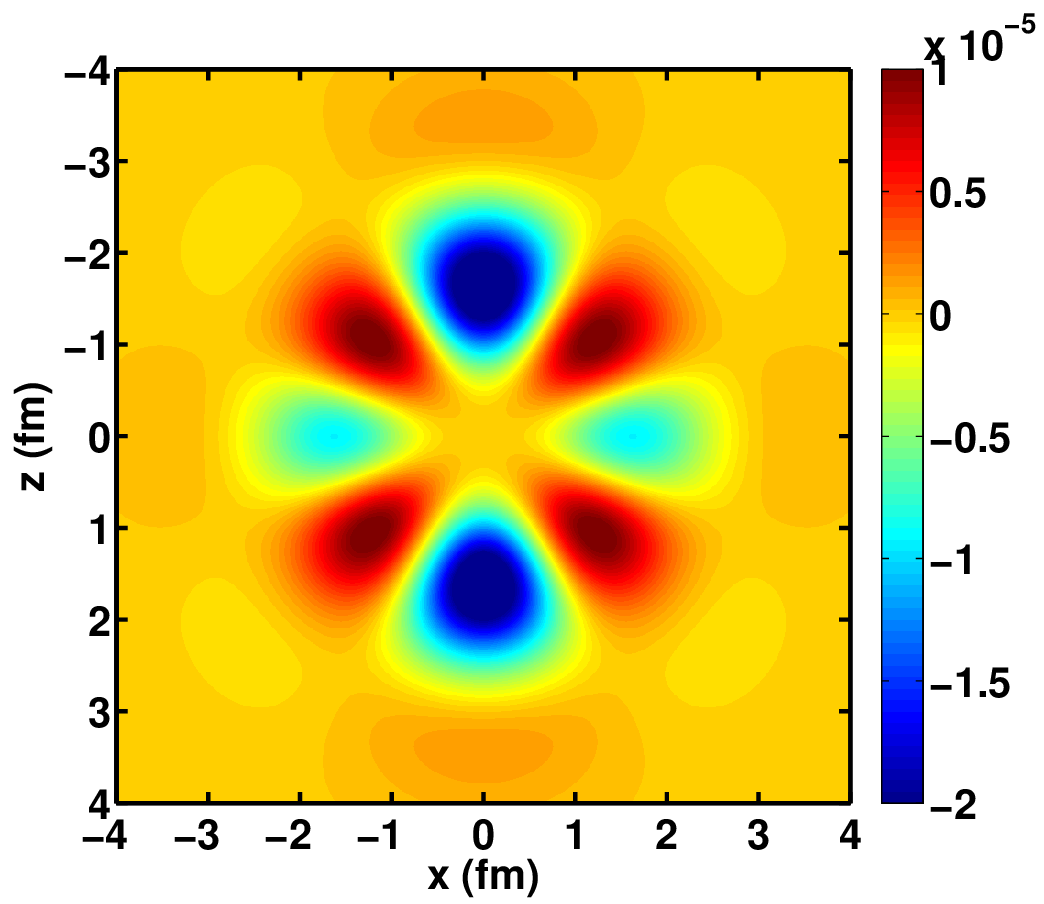}
\qquad\includegraphics[width=0.95\columnwidth]{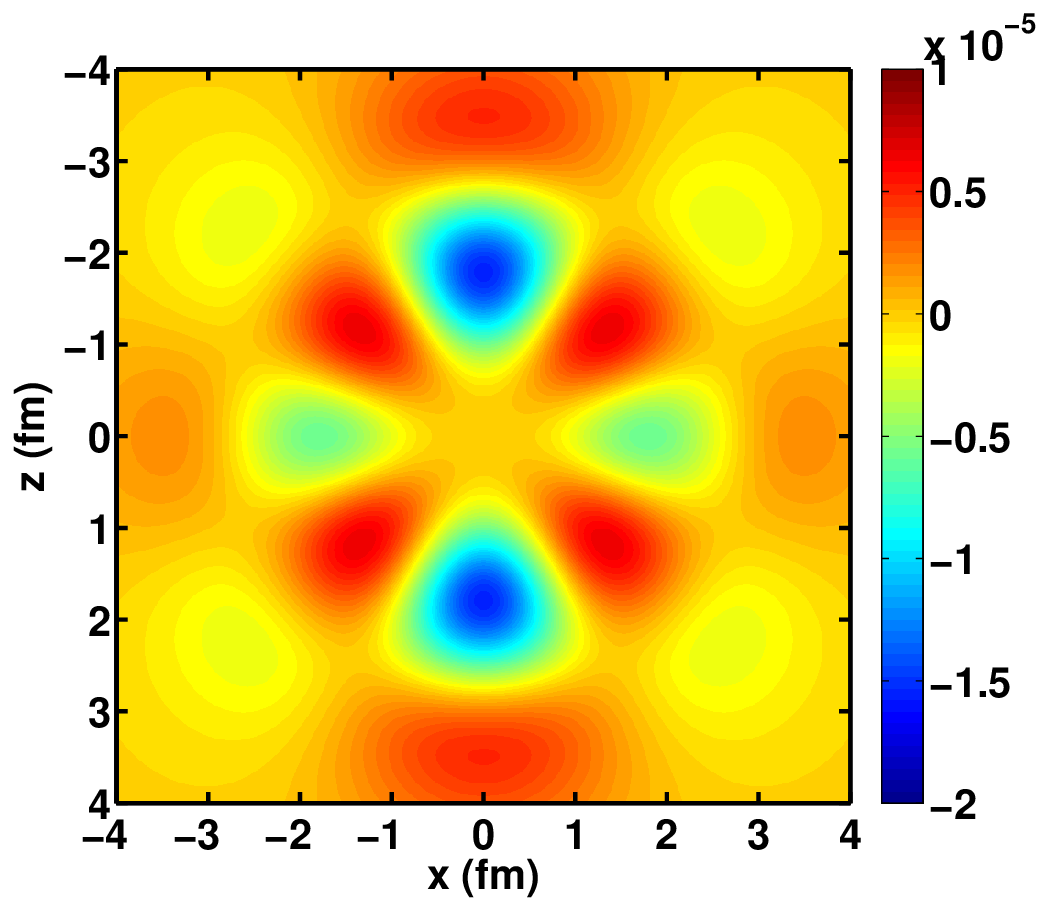}
\caption{(Color online)
  The $y=0$ slices of the translationally-invariant proton
  (left) and neutron (right) densities of the $2^+$ gs of $^8$Li.
  From top to bottom, we present the monopole, quadrupole and hexadecapole
  densities respectively.  These densities were
  calculated at $N_{\max}=12$ and $\hbar\Omega=12.5$~MeV.
  \label{Fig:dens8Li_new}}
\end{figure*}

Another way of visualizing these ti densities is by plotting their
multipole components, $\rho^{(K)}(r)$ 
(we omit the ``ti" subscript for compactness of notation).
In order to exhibit their long distance features, we present the 
magnitudes of the $\rho^{(K)}(r)$ in 
Fig.~\ref{Fig:dens8LiMultiPole} for the $2^+$ gs and the first excited
$4^+$ state of $^8$Li as semilog plots out to a radius of 6 fm.  To
illustrate the radial and angular nodal structure together out to 4 fm, we present
in Fig.~\ref{Fig:dens8Li_new} the $y=0$ slices of the 
multipole components for the gs proton density (left panels) and gs neutron density 
(right panels) defined as their full contribution to the total density. That is,
Fig.~\ref{Fig:dens8Li_new} displays slices of the respective 
full terms contributing to the sum given in Eq.~(\ref{Eq:multipoleti}).  
Hence, the sum of the proton (neutron) multipoles in Fig.~\ref{Fig:dens8Li_new} 
produces the top left (top right) panel of Fig.~\ref{Fig:dens8Li} out to 4 fm.

Let us consider Fig.~\ref{Fig:dens8LiMultiPole} in more detail.
Qualitatively, the multipole components look
very similar for the gs protons and gs neutrons.  The
main difference seems to be that the proton densities fall off more
rapidly with $r$ than the neutron densities for all three multipole components.  
This is understandable
since this is a neutron-rich system and the single-neutron removal
energy is less than that of the single-proton removal energy.
  Note however that the clustering of the neutrons in the
gs of $^8$Li  (left panel of Fig.~\ref{Fig:dens8LiZoom}) is not
evident from the multipole components $\rho^{(K)}$ of the neutron
density displayed in the top right panel of 
Fig.~\ref{Fig:dens8LiMultiPole}.  That is, even though the radial
multipole densities of the protons and neutrons look qualitatively
similar, the corresponding 3-dimensional densities look qualitatively
different.   
Apart from overall difference in the asymptotic behavior, 
the biggest difference between the proton and neutron multipole densities 
is in the interior of the nucleus, below 2.5  fm, 
where the magnitude of the proton hexadecapole density is up to 60\% 
larger than that of the neutron hexadecapole density.  
On the other hand, in the exterior region, beyond 4 fm, 
the neutron hexadecapole density is more than 
an order of magnitude larger than the proton density.

The monopole proton and neutron densities of the first excited $4^+$ state of
$^8$Li seen in Fig.~\ref{Fig:dens8LiMultiPole} are similar to 
those of the ground state, with the proton
density falling off more rapidly than the neutron density.  On the
other hand, the higher multipole components $\rho^{(K)}(r)$ of the
first excited $4^+$ state of $^8$Li look qualitatively quite
different than those of the ground state. In addition, the
protons and the neutrons of the excited $4^+$ state differ from each 
other in their higher multipole components.  The quadrupole density of the neutrons in
the $4^+$ state has a node, in contrast to that of the protons.
Neither the proton nor the neutron hexadecapole density has a node in the
$4^+$ state, whereas both the proton and the neutron hexadecapole densities
have a node in the ground state.  

We turn our attention now to the nodal structures of the multipole components of
the charge-dependent density distributions for the gs of $^8$Li in Fig.~\ref{Fig:dens8Li_new}.  
Since each panel represents a term contributing to Eq.(~\ref{Eq:multipoleti}), we observe 
the angular nodal structure governed by the respective spherical harmonic factor.  
The radial nodes arise from the radial functions $\rho^{(K)}(r)$ displayed in 
Fig.~\ref{Fig:dens8LiMultiPole}.
The significant differences between the proton and the neutron magnitudes at each multipolarity
reflect the differences in the magnitudes of the radial distributions of the multipoles 
seen above in the top two panels of Fig.~\ref{Fig:dens8LiMultiPole}.

From both Figs.~\ref{Fig:dens8LiMultiPole} and \ref{Fig:dens8Li_new}, one observes that the gs proton monopole density has a slightly higher central value than the neutron gs monopole density, but it falls off much more rapidly with $r$ than the neutron density.  On the other hand, the gs neutron quadrupole density has larger (in magnitude) features than the gs proton density, both at small and large distances.   And again, the gs hexadecapole densities have the most interesting features, including the nodal structure at about 3 fm.   In the interior region of the nucleus, the gs proton hexadecapole density has more pronounced features, whereas in region beyond 3 fm the gs neutron hexadecapole density has more pronounced features.   However, none of these multipole density plots show any hint of clustering of the neutrons; nevertheless, the plots of the three gs neutron multipole densities of Fig.~\ref{Fig:dens8Li_new} add up to give the total translationally-invariant density of the left panel of 
Fig.~\ref{Fig:dens8LiZoom}, which does indicate two neutron clusters, 
with their centers separated by about 1.5 fm.

Although these densities are not (yet) fully converged, we
feel that the qualitative features will persist in the limit of a
complete basis.  In particular salient differences between different
states and/or between the proton and neutron densities are likely to
survive in that limit.

\section{Summary and Outlook}

We have performed no-core full configuration calculations for the
Lithium isotopes, $^6$Li, $^7$Li, and $^8$Li with the realistic NN
interaction JISP16. Several observables obtained (gs energies,
excitation energies, magnetic dipole moments and reduced magnetic dipole
transition probabilities) compare well with both experiment and alternate
methods and interactions.  For certain observables that are more
sensitive to long-range correlations (the RMS radius, electric
quadrupole moments, and reduced quadrupole transition probabilities) we were
unable to obtain full convergence, though they also compare favorably
with alternate methods and interactions.

One-dimensional and three-dimensional translationally-invariant
one-body density distributions were calculated for various ground and
excited states of $^6$Li, $^7$Li, and $^8$Li.  These one-body density
distributions provide an excellent framework for visualization of
nuclear shape distortions and clustering effects.  The associated
one-body density matrix in the HO basis provides a compact form of all
of the quantum one-body information for a given nuclear state.

To improve our convergence, especially for matrix elements of
long-range operators, we would require significant increases in basis
space sizes (increased $N_{\max}$) and/or alternatives to the HO
single-particle basis.  Recent advances in the ``Importance-Truncated
No-Core Shell Model"~\cite{Roth:2007sv,Roth:2009cw}, the
``Symmetry-Adapted No-Core Shell Model"~\cite{Dytrych:2008zz} 
and the ``No-Core Monte Carlo Shell Model" \cite{Abe2011,Abe2012}
are promising new methods for accessing much larger basis spaces.

Further advances in NN interactions, as well as three-body forces,
could also help resolve some of the residual differences between
theory and experiment.  Of course, there is also the possibility that
four-body forces may play a significant role.

\begin{acknowledgments}
We would like to thank Mark Caprio, Michael Kruse, John Millener, and
Petr Navr\'atil for stimulating discussions.  This work was supported in
part by U.S. Department of Energy Grant DE-FC02-09ER41582
(SciDAC/UNEDF) and DE-FG02-87ER40371, and by the US NSF grant 0904782.
Computational resources were provided by the National Energy Research
Supercomputer Center (NERSC), which is supported by the Office of
Science of the U.S. Department of Energy.
\end{acknowledgments}

\appendix

\section{Deconvolution of the one-body density}

In our many-body framework, we use the Slater determinants of a
single-particle Harmonic Oscillator (HO) basis.  In this basis, we can
define a space-fixed One-Body Density Matrix (OBDM) by its matrix
elements
\begin{eqnarray}
 \rho_{\beta\alpha}^{fi} &=& 
  \langle \Psi_f | a^\dagger_\alpha a_\beta | \Psi_i \rangle \,,
\label{Eq:OBDME_HO}
\end{eqnarray}
where $\alpha$ and $\beta$ stand for a set of single-particle quantum
numbers $(n_\alpha, l_\alpha, j_\alpha, m_\alpha, \tau_{z,\alpha})$
and $(n_\beta, l_\beta, j_\beta, m_\beta, \tau_{z,\beta})$.  These
matrix elements $\rho_{\beta\alpha}^{fi}$, together with the
expressions for the single-particle wavefunctions
$\psi_{\alpha}(\vec{r})$, completely determine the OBDM in coordinate
space
\begin{eqnarray}
 \rho^{fi}(\vec{r},\vec{r}') &=& \sum_{\alpha, \beta} 
      \rho_{\beta\alpha}^{fi} \;
      \psi^\star_{\alpha}(\vec{r}) \; \psi_{\beta}(\vec{r}') \,.
\end{eqnarray}
The local one-body density becomes
\begin{eqnarray}
 \rho_{\rm sf}^{\Omega}(\vec{r}) &=& \sum_{\alpha, \beta} 
      \rho_{\beta\alpha} \;
      \psi^\star_\alpha(\vec{r}) \; \psi_\beta(\vec{r}) \,.
\end{eqnarray}
where
\begin{eqnarray}
 \psi_\alpha(\vec{r})&=&<\vec r | \alpha > \nonumber\\
 &=& \sum_{m_l,m_s}\langle l_\alpha m_{l} s_\alpha m_{s}|j_\alpha m_{j\alpha}\rangle
\phi_{n_\alpha l_\alpha m_{l}}(\vec{r})\nonumber\\&&\times\chi_{s_\alpha m_s}
\end{eqnarray}
with $\phi_{n_\alpha l_\alpha m_{l}}(\vec{r})$ as the usual three-dimensional HO 
coordinate-space wavefunction and $\chi_{s_\alpha m_s}$ is the Pauli spinor for the nucleon.  While we retain the same basis functions $\psi_\alpha$ for neutrons and for protons (subject to potentially different single-particle state cutoffs as dictated by $N_{\rm max}$), our many-body states treat the neutrons and protons independently so isospin is not a conserved quantity.
The one-body density distribution is normalized to the number of nucleons
\begin{eqnarray}
 \int\rho_{\rm sf}^{\Omega}(\vec{r}) d^3r &=& A \,.
\end{eqnarray}

This local one-body density distribution includes contributions from
the cm (center of mass) motion of the many-body wavefunctions
$\Psi(\vec r_1.....\vec{r}_A)$, hence the subscript sf (space-fixed).
However, because of the exact factorization of the cm wavefunction and
the ti (translationally-invariant) wavefunction, the sf density can be
expressed as a convolution of the ti density distribution $\rho_{\rm
  ti}$ with the cm density distribution $\rho_{\rm cm}$ via
\begin{eqnarray}
  \rho_{\rm sf}^{\Omega}(\vec{r}) &=& 
  \int\, \rho_{\rm ti}(\vec{r}-\vec{R}) \; \rho_{\rm cm}^{\Omega}(\vec{R})\; d^3\vec{R} \,.
\end{eqnarray}
For a HO basis, $\rho_{\rm cm}^{\Omega}$ is a simple Gaussian (the gs density
of $H_{\rm cm}$) that smears out any non-spherical features of
$\rho_{\rm ti}$.  This smearing also introduces into $\rho$ an undesired
dependence on the basis parameter $\hbar\Omega$ that could mask the
convergence of $\rho_{\rm ti}$. Given our expansion in a HO basis, we have analytically evaluated Eq. (A7) using our OBDMs with \emph {Mathematica}~\cite{Wolfram}.  
This produces an analytic expression for the ti OBDM. 

In order to display properties of the density $\rho_{\rm ti}$, as well as
investigate its convergence, we deconvolute the spurious cm density
using well-known Fourier methods~\cite{Morse}.  The ti density is
given by
\begin{eqnarray}
 \rho_{\rm ti}(\vec{r}) &=& 
  F^{-1}\bigg[\frac{F[\rho_{\rm sf}^{\Omega}(\vec{r})]}{F[\rho_{\rm cm}^{\Omega}(\vec{R})]}\bigg]
\end{eqnarray}
where $F[f(\vec r)]$ is the 3-dimensional Fourier transform of $f(\vec r)$.

For an $A$-nucleon eigenstate with total angular momentum $J$ and
projection $M$, and possibly additional quantum numbers $\lambda$,
denoted by $|A \lambda J M\rangle$, we can evaluate the local
space-fixed density by evaluating the matrix element~\cite{AngMom}
(henceforward, we omit the superscript "$\Omega$" for compactness of notation)
\begin{eqnarray}
 \rho_{\rm sf}(\vec{r}) &=&
\langle A\lambda J M |\hat{\rho}_{\rm sf}(\vec{r})|A\lambda J M\rangle 
\end{eqnarray}
using the local density operator
\begin{eqnarray}
 {\hat\rho}_{\rm sf}(\vec{r}) &=& \sum_{k=1}^A\delta^3(\vec{r}-\vec{r}_k)\nonumber\\
  &=& \sum_{k=1}^A\frac{\delta(r-r_k)}{r^2}
  \; \sum_{lm}Y^{\star m}_{l}(\hat{r}_k)Y^m_{l}(\hat{r})
\end{eqnarray}
where $\hat{r}$ is the unit vector in the direction $\vec{r}$, and
$Y^m_l(\hat{r})$ is a spherical harmonic.  Note that it has the
property
\begin{eqnarray}
 Y^{-m}_l(\hat{r})&=&(-1)^mY_l^{\star m}(\hat{r}) \,.
\end{eqnarray}
To efficiently perform the deconvolution of the spurious cm density
and the ti density, we make a multipole expansion of the
local density~\cite{AngMom}
\begin{eqnarray}
 \rho_{\rm sf}(\vec{r}) &=&
\sum_K \frac{\langle J M K 0|J M\rangle}{\sqrt{2J+1}}  
  \; Y_{K}^{\star 0}(\hat{r}) \; \rho_{\rm sf}^{(K)}(r) 
\label{Eq:multipole}
\end{eqnarray}
where $\rho_{\rm sf}^{(K)}(r)$ is the $K^{th}$ multipole of the sf
density.  The same equation with``sf" replaced by ``ti" provides 
the corresponding expansion for the local ti density. For initial and 
final states with spin $J$, the multipoles
range from $K=0$ to $K = 2 J$.  This multipole expansion greatly
simplifies the Fourier transforms needed for the deconvolution, as we
will see at the end of this appendix.

With a HO single-particle basis, each multipole is given by
\begin{eqnarray}
\rho_{\rm sf}^{(K)}(r) &=& \sum R_{n_1l_1}(r) R_{n_2l_2}(r) \;
\frac{-1}{\hat{K}} \langle l_1\frac{1}{2}j_1||Y_K||l_2\frac{1}{2}j_2\rangle
\nonumber\\
& &\times\langle A\lambda J ||(a_{n_1l_1j_1}^\dagger
\tilde{a}_{n_2l_2j_2})^{(K)}||A\lambda J \rangle \,,
\label{Eq:rhoK}
\end{eqnarray}
where $\hat{K}=\sqrt{2K+1}$.  The $R_{nl}(r)$'s are the radial
compnents of the HO wavefunction
\begin{eqnarray}
 R_{nl}(r)&=&\bigg[\frac{2(2\nu)^{l+3/2}\Gamma(n+1)}{\Gamma(n+\frac{l}{2}+\frac{3}{2})}\bigg]^{1/2}
 e^{-\nu r^2}L_n^{l+\frac{1}{2}}(2\nu r^2)
\nonumber\\
\end{eqnarray}
with $L_n^{l+\frac{1}{2}}$ the associated Laguerre polynomials 
and $\nu = m c^2 \hbar \Omega/(2 \hbar^2 c^2)$.  The
reduced matrix element of a spherical harmonic in Eq.~(\ref{Eq:rhoK})
can be written as
\begin{eqnarray}
\lefteqn{ \langle l_1\frac{1}{2}j_1||Y_K||l_2\frac{1}{2}j_2\rangle\
\;=\;\frac{1}{\sqrt{4\pi}}\hat{j_1}\hat{j_2}\hat{l_1}\hat{l_2}}
\nonumber\\
&&\times (-1)^{j_1+\frac{1}{2}} \; \langle l_10l_20|K0\rangle \;
\bigg\{ \begin{array}{ccc}
j_1 & j_2 & K\\
l_2 & l_1 & \frac{1}{2}\\
\end{array} \bigg\}
\end{eqnarray}
using a Wigner-6J coefficient.  (We use the Conden--Shortley
convention for the Clebsch--Gordan coefficients, defined in
Ref.~\cite{Talmi}.)

Finally, $\langle A\lambda J ||(a_{n_1l_1j_1}^\dagger
\tilde{a}_{n_2l_2j_2})^{(K)}||A\lambda J\rangle$ in
Eq.~(\ref{Eq:rhoK}) represents a reduced matrix element of the
$K^{th}$ multipole of the OBDM.  The $K^{th}$ multipole operator for
initial and final states with the same $M_j$ can be written as
\begin{eqnarray}
 (a_{n_1l_1j_1}^\dagger \tilde{a}_{n_2l_2j_2})^{(K)} &=&\sum_{m_j}
(-1)^{j_2+m_j} \, \langle j_1 m_j j_2 -m_j|K0\rangle 
\nonumber\\
&& {} \times  a^\dagger_{n_1 l_1 j_1 m_j}a_{n_2 l_2 j_2 m_j} \,.
\end{eqnarray}
Note that in Eq.~(\ref{Eq:rhoK}) we use $M_j$-independent reduced
matrix elements, defined by the Wigner--Eckart theorem.  For a generic
operator $T_{Kk}$ the reduced matrix element is defined by
\begin{eqnarray}
\langle\lambda_fJ_f||T_K||\lambda_iJ_i\rangle &=& 
 \hat{J}_f \frac{\langle\lambda_fJ_fM_f|T_{Kk}|\lambda_iJ_iM_i\rangle}
 {\langle J_fM_fKk|J_iM_i\rangle}
\nonumber \\
\end{eqnarray}
provided that the Clebsch-Gordan coefficient in the denominator is not
zero.  Thus the reduced matrix elements of the $K^{th}$ multipole of
the OBDM are given by
\begin{eqnarray}
\lefteqn{ \langle A\lambda J ||(a_{n_1l_1j_1}^\dagger \tilde{a}_{n_2l_2j_2})^{(K)}||A\lambda J\rangle
\;=\;}
\nonumber \\
&=& \frac{\sqrt{2J+1}}{\langle J M K 0|J M \rangle}
\sum_{m_j} (-1)^{j_2+m_j} \, \langle j_1 m_j j_2 -m_j|K0\rangle 
\nonumber\\
&& {} \times  \langle A\lambda J M | a^\dagger_{n_1 l_1 j_1 m_j}a_{n_2 l_2 j_2 m_j} |A\lambda J M\rangle \,.
\end{eqnarray}
where $\langle A\lambda J M | a^\dagger_{n_1 l_1 j_1 m_j}a_{n_2 l_2 j_2 m_j} |A\lambda J M\rangle$ 
are the OBDME's in HO space defined in Eq.~(\ref{Eq:OBDME_HO}).  Note
that the factor $\sqrt{2J+1}/{\langle J M K 0|J M \rangle}$ in this
expression cancels against the same factor in the expression for
$\rho_{\rm sf}(\vec{r})$, Eq.~(\ref{Eq:multipole}).

We can now efficiently perform the deconvolution, by using the Fourier
transform properties of the multipole expansion of the local density
\begin{eqnarray}
\int d^3\vec{r} \exp(i\vec{q}\cdot\vec{r})
 \; \rho_{\rm sf}^{(K)}(r) \; i^KY_K^{\star 0}(\hat{r}) &=&
 \tilde{\rho}_{\rm sf}^{(K)}(q) \; Y_K^{\star 0}(\hat{q}) \,,
\nonumber \\
\end{eqnarray}
where the multipole component of the density in momentum space is expressed as
\begin{eqnarray}
 \tilde{\rho}_{\rm sf}^{(K)}(q) &=& 4\pi\int j_K(qr)\; \rho_{\rm sf}^{(K)}(r) \; r^2dr
\end{eqnarray}
with $j_K$ the spherical Bessel Functions of the first kind.

Thus the deconvolution of each multipole gives
\begin{eqnarray}
  \rho_{\rm ti}^{(K)}(r) 
&=& 
  \frac{1}{2\pi^2}
  \int j_K(qr)\; 
  \frac{\tilde{\rho}_{\rm sf}^{(K)}(q)}{\tilde{\rho}_{\rm cm}(q)}
  \; q^2dq
\end{eqnarray}
where
\begin{eqnarray}
 \tilde{\rho}_{\rm cm}(\vec{q}) &=& \frac{\tilde{\rho}_{\rm cm}^{(0)}(q)}{2\sqrt{\pi}}
\nonumber\\
&=& \frac{8\sqrt{2}}{\sqrt{\pi}} \, \nu^{3/2} \int_0^\infty
  \frac{e^{-2\nu R^2}\sin(qR)}{qR} \; R^2dR \nonumber \\
&=& e^{ -q^2/8\nu}
\end{eqnarray}

For spherically symmetric nuclei, this
deconvolution simplifies even further because we only have one term in
the multipole expansion, $K=0$
\begin{eqnarray}
 \rho_{\rm ti}^{(0)}(r) &=& 
 \frac{1}{2\pi^2}\int_0^\infty \frac{\sin(qr)}{qr} \; 
 \frac{\tilde{\rho}_{\rm sf}^{(0)}(q)}{\tilde{\rho}_{\rm cm}(q)} \; q^2dq \,,
\end{eqnarray}
and the 3-dimensional ti density is simply
\begin{eqnarray}
 \rho_{\rm ti}(\vec{r}) &=& \frac{\rho_{\rm ti}^{(0)}(r)}{2\sqrt{\pi}}  
\end{eqnarray}
without any angular dependence.

Another advantage of the multipole expansion is that it allows for a
straightforward calculation of the (sf or ti) density for any magnetic
projection $M$, once the multipoles $\rho^{(K)}(r)$ are known.  The
multipoles $\rho^{(K)}(r)$ are completely determined from reduced matrix
elements, wich do not depend $M$.  The only dependence of 
$\rho_{\rm  sf}(\vec{r})$ on $M$ is entirely through the explicitly
$M$-dependent Clebsch--Gordan coefficients in
Eq.~(\ref{Eq:multipole}).

Finally, the expressions here are specific for the local static
density (the same final and initial state), but the extension to
local transition densities is straightforward.


\end{document}